\documentclass[aps,twocolumn,showpacs,preprintnumbers,amsmath,amssymb,prb,nofootinbib,10pt]{revtex4-1}
\pdfoutput=1
\usepackage{graphicx}
\usepackage{dcolumn}
\usepackage[tight]{subfigure}
\usepackage{amsmath}
\usepackage{verbatim}
\usepackage{color}
\usepackage{bm} 
\usepackage{bbm}
\usepackage{mathbbol}
\usepackage{natbib}
\usepackage{xspace}
\usepackage{marginnote}
\usepackage{mathtools}
\usepackage{dsfont}
\usepackage{hyperref} \hypersetup{colorlinks=true,linktoc=all,linkcolor=blue,breaklinks=true,citecolor=blue,urlcolor=blue}
\usepackage{etoolbox}
\usepackage[normalem]{ulem}
\usepackage[bottom]{footmisc}

\robustify{\uparrow}
\robustify{\downarrow}
\robustify{\sum}
\robustify{\int}
\robustify{\nonumber}
\robustify{\cite}
\robustify{\footnote}
\newrobustcmd{\Figure}[2]{
  \begin{figure}[ht]
    \includegraphics[width=1.0\linewidth]{#1}
    \caption{#2}
  \end{figure}
}
\expandafter\newrobustcmd\csname Figure2\endcsname[3]{
\begin{figure}[ht]
  \includegraphics[width=0.9\linewidth]{#1}
  \\
  \includegraphics[width=0.9\linewidth]{#2}
  \caption{#3}
\end{figure}
}
\renewrobustcmd{\Re}{{\text{Re}}}
\renewrobustcmd{\Im}{{\text{Im}}}
\newrobustcmd{\one}{\mathds{1}}
\DeclareMathOperator{\Tr}{Tr}
\newrobustcmd{\Eq}[1]{Eq.~(\ref{#1})}
\newrobustcmd{\Eqs}[1]{Eqs.~(\ref{#1})}
\newrobustcmd{\BrackEq}[1]{[Eq.~(\ref{#1})]}
\newrobustcmd{\BrackApp}[1]{[App.~(\ref{#1})]}
\newrobustcmd{\BrackSec}[1]{[Sec.~(\ref{#1})]}
\newrobustcmd{\eq}[1]{(\ref{#1})}
\newrobustcmd{\Fig}[1]{Fig.~\ref{#1}}
\newrobustcmd{\BrackFig}[1]{[Fig.~\ref{#1}]}
\newrobustcmd{\fig}[1]{\ref{#1}}
\newrobustcmd{\Figs}[1]{Figs.~\ref{#1}}
\newrobustcmd{\Sec}[1]{Sec.~\ref{#1}}
\newrobustcmd{\Ref}[1]{Ref.~\onlinecite{#1}}
\newrobustcmd{\Refs}[1]{Refs.~\onlinecite{#1}}
\newrobustcmd{\App}[1]{App.~\ref{#1}}
\newrobustcmd{\fpOp}{(-\one)^N}
\newrobustcmd{\nbrack}[1]{\left(#1\right)}
\newrobustcmd{\sqbrack}[1]{\left[#1\right]}
\newrobustcmd{\cbrack}[1]{\left\{#1\right\}}
\newrobustcmd{\nlbrack}[1]{\left(#1\right.}
\newrobustcmd{\sqlbrack}[1]{\left[#1\right.}
\newrobustcmd{\clbrack}[1]{\left\{#1\right.}
\newrobustcmd{\nrbrack}[1]{\left.#1\right)}
\newrobustcmd{\sqrbrack}[1]{\left.#1\right]}
\newrobustcmd{\crbrack}[1]{\left.#1\right\}}
\newrobustcmd{\cosfn}[1]{\cos{\left(#1\right)}}
\newrobustcmd{\sinfn}[1]{\sin{\left(#1\right)}}
\newrobustcmd{\tanhfn}[1]{\tanh\!\left(#1\right)}
\newrobustcmd{\phiL}{\phi_{\text{L}}}
\newrobustcmd{\phiR}{\phi_{\text{R}}}
\newrobustcmd{\meff}{m^*_{e}}
\newrobustcmd{\gamL}{\gamma_{\text{L}}}
\newrobustcmd{\gamR}{\gamma_{\text{R}}}
\newrobustcmd{\Htoy}{H_{\text{t}}}
\newrobustcmd{\ttoy}{t_{\text{t}}}
\newrobustcmd{\Jtoy}{J_{\text{t}}}
\newrobustcmd{\Hc}{\text{H.c.}}
\newrobustcmd{\delphi}{\delta\phi}
\newrobustcmd{\Itoy}{I_{\text{t}}}
\newrobustcmd{\ptoy}{p_{\text{t}}}
\newrobustcmd{\Btoy}{B_{\text{t}}}
\newrobustcmd{\dtoy}{d_{\text{t}}}
\newrobustcmd{\alphaToy}{\alpha_{\text{t}}}
\newrobustcmd{\eqavg}[1]{\left\langle#1\right\rangle_{\text{eq}}}
\newrobustcmd{\fM}{f_{\text{M}}}
\newrobustcmd{\fMdag}{f^\dagger_{\text{M}}}
\newrobustcmd{\nM}{n_{\text{M}}}
\newrobustcmd{\pM}{p_{\text{M}}}
\newrobustcmd{\unit}[1]{\text{#1}}
\newrobustcmd{\pGS}{p_{\text{GS}}}
\newrobustcmd{\lB}{l_{\text{b}}}
\newrobustcmd{\VB}{V_{\text{b}}}
\newrobustcmd{\lW}{l_{\text{w}}}
\begin{document}
%
% Title
%
\title{Absence of supercurrent sign reversal in a topological junction with a quantum dot
}
\author{J. Schulenborg}
\author{K. Flensberg}
\affiliation{
  Center for Quantum Devices, Niels Bohr Institute, University of Copenhagen, 2100 Copenhagen, Denmark
}
% \pacs{
%   85.75.-d,
%   73.63.Kv,
%   85.35.-p
% }
%
% Abstract
%
\begin{abstract}
Experimental techniques to verify Majoranas are of current interest. A prominent test is the effect of Majoranas on the Josephson current between two wires linked via a normal junction. Here, we study the case of a quantum dot connecting the two superconductors and the sign of the supercurrent in the trivial and topological regimes under grand-canonical equilibrium conditions, explicitly allowing for parity changes due to, e.g., quasi-particle poisoning. We find that the well-known supercurrent reversal for odd occupancy of the quantum dot ($\pi$-junction) in the trivial case does not occur in the presence of Majoranas in the wires. However, we also find this to be a mere consequence of Majoranas being zero energy states. Therefore, the lack of supercurrent sign reversal can also be caused by trivial bound states, and is thus not a discriminating signature of Majoranas.
\end{abstract}
\maketitle

\section{Introduction}
\label{sec_introduction}

Majorana bound states in condensed-matter systems have attracted enormous interest in the last decade~\cite{Nayak2008Sep,Wilczek2009Sep,Franz2010Mar,Stern2010Mar,Leijnse2012Nov}, owing mostly to their fundamental, quantum statistical properties and their potential application in fault-tolerant, topological quantum computing~\cite{Kitaev2003Jan,Nayak2008Sep,Sarma2015Oct}. A promising candidate among several potential physical systems~\cite{Kitaev2003Jan,Fu2008Mar,Lutchyn2010Aug,Oreg2010Oct,Cook2011Nov,Choy2011Nov,Vazifeh2013Nov,Klinovaja2013Nov,Beenakker2013Mar,Nadj-Perge2014Oct,Elliott2015Feb,Lutchyn2018May,Fornieri2019Apr} to host such states are semiconducting, quasi one-dimensional nanowires with proximity-induced s-wave superconductivity, Rashba spin-orbit coupling as well as a specifically tuned parallel magnetic field~\cite{Kitaev2001,Lutchyn2010Aug,Oreg2010Oct,Mourik2012May}. 
However, despite considerable experimental evidence for the presence of Majoranas in such wires~\cite{Mourik2012May,Das2012Nov,Rokhinson2012Sep,Wiedenmann2016Jan,Deng2016Dec,Aguado2017Oct,Zhang2018Mar,Lutchyn2018May}, unambiguously distinguishing Majoranas from regular Andreev subgap states~\cite{Akhmerov2011Jan,Das2012Nov,Lee2013Dec,Cayao2015Jan,San-Jose2016Feb,Liu2017Aug,Liu2018Jun,Hell2018Apr,Vuik2019Nov,Chiu2019Jan} remains challenging.

In principle, a desired way to unambiguously trace and exploit Majoranas in solid state systems is to measure and manipulate observables directly affected by the most distinctive properties of Majoranas --- their statistics and their insusceptibility to local decoherence~\cite{Kitaev2003Jan,Nayak2008Sep,Lutchyn2018May}. Two key long-term goals of this line of research are the possibility of braiding~\cite{Read2000Apr,Ivanov2001Jan,Kitaev2003Jan,Alicea2011Feb,Flensberg2011Mar,Sau2011Sep,vanHeck2012Mar,Aasen2016Aug} and the successful implementation of fault-tolerant, Majorana-based qubits~\cite{Sarma2015Oct,Aasen2016Aug,Karzig2017Jun,Plugge2017Jan,Litinski2017Sep}. However, one major challenge is that building and operating such devices needs many components and fine tuning. 
It is thus important to find more easily applicable test criteria that, while not always fully conclusive, still provide sufficient confidence to further advance in conceiving and building the final device.

%%%%%%%%%%%%%%%%%%%%%%%%%%%%%%%%%%%%%%%%%%%%%%%%%%%%%%%%%%%%%
\begin{figure}
	\centering
	\includegraphics[width=\linewidth]{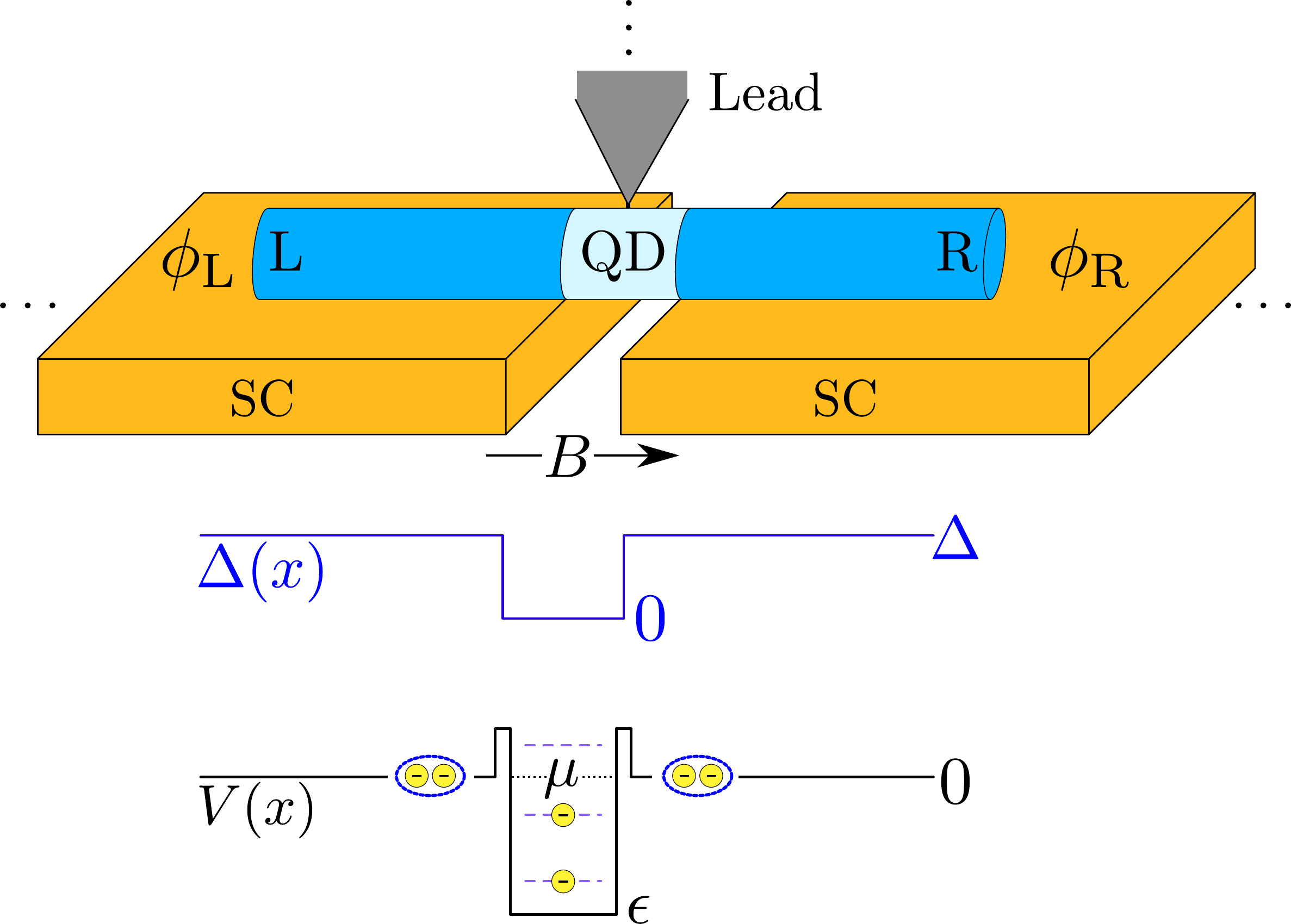}
	\caption{The system of interest consists of two one-dimensional, equally long (length $\lW$) nanowires L and R that feature Rashba spin-orbit coupling and proximity-induced s-wave superconductivity, and of a quantum dot QD coupling the two wires. The left and right superconducting phases are denoted by $\phiL,\phiR$, the blue line sketches the superconducting gap $\Delta(x) > 0$, assuming the constant $\Delta > 0$ in the wires and disappearing in the dot. The potential landscape $V(x)$ is indicated by the black line. In the wires, $V(x)$ coincides with the equilibrium chemical potential $\mu = 0$; in the dot, it is lowered considerably to $\epsilon < 0$ with $|\epsilon|/\Delta \gg 1$. The square-potential barriers overlap half with the wire regions $(\Delta(x) > 0)$, and half with the dot region $(\Delta(x) = 0)$; they have finite length $\lB$ and height $\delta\VB > 0$ compared to $\mu = 0$. 
	The quantum dot furthermore couples to a non-superconducting lead functioning as a probe of the density of states in the wire-dot-wire system.
	}
	\label{fig_setup}
\end{figure}
%%%%%%%%%%%%%%%%%%%%%%%%%%%%%%%%%%%%%%%%%%%%%%%%%%%%%%%%%%%%%

First experiments aiming to find zero-energy Majorana states in nanowires focussed on measuring the corresponding zero-bias conductance peak~\cite{Mourik2012May,Deng2012Dec,Liu2012Dec}. Since the nanowire-based devices of interest are typically integrated into electrical circuits anyhow, such measurements have since become a relatively straightforward consistency check for the existence of Majoranas. Yet, due to the many possible causes of zero-bias peaks including trivial zero-energy Andreev bound states, they are far from conclusive. Alternatively, one can measure the $4\pi$-phase-periodicity in the Josephson current between two topological, effectively p-wave superconducting wires separated by a semi-conducting interface~\cite{Kitaev2001,Kwon2004Feb,vanHeck2011Nov,Law2011Aug,Rokhinson2012Sep,Wiedenmann2016Jan,Stefanski2016Oct,Chiu2019Jan}. However, just as for zero-bias peaks, the $4\pi$-periodicity is only a necessary, but not sufficient condition for Majoranas~\cite{Chiu2019Jan}. Furthermore, experimental issues arise from the critical dependence on the conservation of parity, and thus on the absence of quasi-particle poisoning during the phase sweep. While difficult under typical equilibrium conditions, measurment of the response to a phase changing on time scales faster than the typical poisoning time~\cite{Rokhinson2012Sep,Wiedenmann2016Jan,Laroche2019Jan} requires that the $4\pi$-periodicity is not caused by Landau-Zener transitions between topologically trivial subgap states. Finally, most recent advances into entropy measurements of nanoscale systems~\cite{Hartman2018Aug} promise to allow~\cite{Sela2019Oct} to distinguish regular zero-energy bound states from Majoranas by accessing their fractional entropy~\cite{Cooper2009Apr,Hou2012Oct,Smirnov2015Nov,Sela2019Oct}, $\Delta S =  (k_\text{B}/2)\ln(2)$.

From a practical point of view, finding evidence of Majoranas in devices consisting of several nanowires that couple via quantum dots or Coulomb islands is of particular interest~\cite{Liu2011Nov,Leijnse2011Oct,Cao2012Sep,Lee2013Jun,Vernek2014Apr,Ruiz-Tijerina2015Mar,Stefanski2016Oct,Deng2016Dec,Albrecht2016Mar,Hoffman2017Jul,Liu2017Aug,Prada2017Aug,Clarke2017Nov,Chevallier2018Jan,Deng2018Aug}, as such devices form the basic building blocks in many braiding and quantum-computation related proposals ~\cite{Flensberg2011Mar,Terhal2012Jun,Sau2012Jul,Aasen2016Aug,Karzig2017Jun,Plugge2017Jan,Litinski2017Sep}. 
Measurable signatures of Majoranas hybridizing with the dot/junction states in these setups are both in the subgap spectra~\cite{Prada2017Aug,Cayao2018May,O'Farrell2018Dec,Deng2018Aug,Ricco2019Apr,Yavilberg2019Dec} and the equilibrium supercurrent~\cite{Camjayi2017Jul,Cayao2017Nov,Cayao2018May,Schrade2018Sep,Ridderbos2018Nov,Awoga2019Sep}. 
Our main focus here is on whether and how the Majoranas interfere with the formation of so-called $\pi$-junctions~\cite{Spivak1991Feb,Rozhkov2001Nov,vanDam2006Aug}, that is, the supercurrent sign reversal concomitant with parity flips in a quantum dot that connects two superconducting wires as in \Fig{fig_setup}.
Specifically, it was suggested that a notable \emph{absence} of such sign changes could be a signature of Majoranas in the wires~\cite{Camjayi2017Jul,Schrade2018Sep,Awoga2019Sep}, both in the absence and presence of quasi-particle poisoning, including the case of conserved time reversal symmetry~\cite{Camjayi2017Jul}.

In this paper, we more closely examine the case of a $B$-field induced, time-reversal symmetry breaking topological phase in grand-canonical equilibrium~\cite{Awoga2019Sep}, explicitly accounting for quasi-particle poisoning.
Performing numerical analyses of the Josephson current through the system displayed in \Fig{fig_setup}, we consider local Coulomb interaction and a tunable potential $\epsilon$ in the dot, theoretically allowing to change the dot occupation one by one for arbitrary superconducting phase differences $\delphi = \phiL - \phiR$ and parallel magnetic fields $B$. We find that step-like sign reversals of the supercurrent (formation of $\pi$-junctions) indeed disappear in the presence of Majoranas. However, as we elucidate with the help of effective low-energy models, this can be solely due to zero-energy wire states hybridizing with the dot, and thereby preventing total parity changes in the set of subgap eigenstates mediating most of the supercurrent.
Topologically trivial subgap states in the wires as discussed in, e.g., \Refs{Liu2018Jun,Hell2018Apr,Vuik2019Nov} could hence have the same effect.

\section{Model and method}
\label{sec_model}
The system of interest is modeled in \Fig{fig_setup}. Two approximately one-dimensional nanowires L and R (length $\lW$) with Rashba spin-orbit coupling and proximity induced s-wave superconductivity are connected by a confined junction of length $l$ that forms a quantum dot (QD). The setup is assumed to be embedded into a superconducting loop, causing a difference in superconducting phase $\phi$ between the left and right side. Furthermore, a magnetic field is applied parallel to the entire wire, here defined to be along the x-axis. The corresponding Bogoluibov-de Gennes Hamiltonian in the basis $(\psi_\uparrow,\psi_\downarrow,\psi^\dagger_\uparrow,\psi^\dagger_\downarrow)$ reads
\begin{equation}
 H = \begin{pmatrix}H_0 & \hat{\Delta}(x) \\ \hat{\Delta}^\dagger(x) & -H_0^* \end{pmatrix},\label{eq_model}
\end{equation}
with  $H_0 =\frac{p_x^2}{2\meff} - \mu + V(x) + B(x)\sigma_x - \alpha\sigma_y p_x$ and $\hat{\Delta}(x) = i\sigma_y\Delta(x) e^{i\phi(x)}$. This includes $p_x = -i\partial_x$, the Pauli matrices $\sigma_{x,y,z}$ in spin space, and the potential $V(x)$ relative to the chemical potential $\mu$. As shown in \Fig{fig_setup}, we assume a constant potential $V(x) = 0$ for $x$ well inside each wire, and $V(x) = \epsilon$ with the potential shift $\epsilon$ in the quantum dot assumed to be tunable  by a gate electrode. At the interfaces to both wires, there are potential barriers of length $\lB$ and height $\delta \VB$. The superconducting gap $\Delta(x)$ is approximated to vanish in the normal region $0 < x < l$ and to be constant, $\Delta > 0$ in both wires. 
The superconducting phase $\phi(x)$ equals $\phiL$ in the left, $0$ in the dot, and $\phiR$ in the right wire, where gauge invariance dictates the dynamics to only depend on $\delta\phi = \phiL - \phiR$. 
The Zeeman energy $B(x)$ assumes the constant $B$ in the wires; for the junction, we add an additional field $\delta B$ to mimic an energy splitting that would be caused by charging energy. The latter is comparable to a Hartree mean-field approximation of the Coulomb interaction~\cite{Hartree1928Jan}. It neglects exchange terms and spin-degeneracy effects at zero magnetic field, but gives the wanted energy splitting that enables to tune the dot electron occupation one by one. Note that we assume $\Delta$ to be large enough compared to the Kondo temperature in order for Kondo correlations between dot and wires to be negligible.
Finally, we take values consistent with recent experiments\cite{O'Farrell2018Dec,Fornieri2019Apr} for typical lengths $\lW,l$, for the effective mass $\meff$, and the spin-orbit coupling $\alpha$. 

Our aim is to extract measurable signatures of Majorana modes present~\cite{Lutchyn2010Aug,Oreg2010Oct} for the model \eq{eq_model} in the topological regime $B > B_C = \sqrt{\Delta^2 + \mu^2}$, extending from the interface with the middle junction into the wires. We consider, on the one hand, the equilibrium \emph{supercurrent} $I = 2\langle\partial_{\delta\phi} H(\delta\phi)\rangle_{\text{eq}}$ flowing from wire to wire due to the phase difference $\delta\phi$, where $\langle\dotsc\rangle_{\text{eq}}$ denotes the average with respect to the grandcanonical ensemble. On the other hand, we study the quasi-particle spectrum of the system as probed by the energy-dependent transmission coefficient $\mathcal{T}(E)$ of the non-superconducting lead coupled to the central junction, see \Fig{fig_setup}. This yields the energy-resolved density of states when the additional broadening and renormalization due to the coupling to the measurement probe as well as to external leads --- here given by the normal lead and the superconductors (SC) \BrackFig{fig_setup} --- is accounted for. Thus, while the spectrum can theoretically be obtained directly from the density of states, our way of explicitly incorporating the probe into the model more closely connects to both previous~\cite{Pillet2010Nov,Chang2013May} and ongoing~\cite{Whiticar2019} experiments.

For the calculation, we second-quantize $H$ \BrackEq{eq_model} and map it to a tight-binding model with $M$ sites by writing $\partial_x$ as a finite difference with spacing $d = (2\lW + l)/(M-1)$. 
Phase rotating the field operators in the wire R and in the central junction by $-\phiR$, we find (positive current means flow of positive charges from left to right)
\begin{equation}
 I = 2\sum_{\sigma=\pm=\uparrow\downarrow}\sqbrack{t\Im\langle c^\dagger_{\text{N},\sigma} c_{\text{L},\sigma}\rangle_{\text{eq}} + \sigma J\Im\langle c^\dagger_{\text{N},\sigma} c_{\text{L},-\sigma}\rangle_{\text{eq}} }.\label{eq_current}
\end{equation}
This includes direct tunneling with amplitude $t = 1/(2\meff d^2)$ and spin-flip tunneling due to the spin-orbit coupling with amplitude $J = \alpha/(2d)$. The symbols $c^\dagger_{\text{N},\sigma},c_{\text{N},\sigma}$ represent the creation and annihilation operators for electrons with spin $\sigma$ on the leftmost site of the central junction, right next to the last site of wire L in which $c^\dagger_{\text{L},\sigma},c_{\text{L},\sigma}$ create and annihilate electrons. The mean-field approximation enables us to calculate the averages $\langle\dotsc\rangle_{\text{eq}}$ numerically using the equation-of-motion method for Matsubara Green's functions, see \App{app_eq}.

The transmission coefficient of the normal lead is obtained by approximating the latter as a 1D chain with a single site weakly and energy-independently coupled to only one site in the center of the normal junction, as suggested by the peaked shape of the lead in \Fig{fig_setup}. The relevant $4\times 4$-subblock $S$ of the scattering matrix is determined by the Mahaux-Weidenm\"uller formula~\cite{Mahaux1968Jun}, see also comment at the end of \App{app_eq}:
\begin{equation}
 S = \one - 2\pi iW^\dagger\sqbrack{E - H + i\pi WW^\dagger}^{-1} W.\label{eq_scattering}
\end{equation}
The couplings ($\delta_{x,y}$ = Kronecker delta)
\begin{equation}
 (W)_{m\eta\sigma,\eta'\sigma'} = -\eta\sqrt{\rho}\delta_{m,\frac{M}{2}}\delta_{\eta,\eta'}\nbrack{t_{\text{n}}\delta_{\sigma\sigma'} - \sigma J_{\text{n}}\delta_{\sigma,-\sigma'}}\label{eq_coupling}
\end{equation}
are determined by the direct and spin-flip tunneling amplitudes $t_{\text{n}}, J_{\text{n}}$ at which the particles$(\eta = +)$/holes$(\eta = -)$ of spin $\sigma = \uparrow\downarrow = \pm$ at site $m$ in the wire-junction-wire system couple to any of the four states associated with particles$(\eta' = +)$/holes$(\eta' = -)$ of spin $\sigma'$ at the first site of the normal lead \BrackFig{fig_setup}. The lead density of states $\rho$ is assumed to be energy-independent (wideband limit). 
The transmission coefficient is finally obtained as 
\begin{equation}
\mathcal{T}(E) = 2 - \Tr\sqbrack{S_{ee}^\dagger S_{ee}} + \Tr\sqbrack{S_{he}^\dagger S_{he}},
\end{equation}
where the $2\times 2$ matrices $S_{ee}$ and $S_{eh}$ contain all elements of $S$ representing electrons from the normal lead reflecting as electrons $(ee)$ or holes $(he)$ back into the lead.

To understand the behavior of $I$ and $\mathcal{T}$, we relate them to the fermion parity $\pGS = \langle\fpOp\rangle_{\text{GS}}$ in the ground state (GS) of $H$, calculated using the Pfaffian as in \Ref{Kitaev2001}. Moreover, we compare the system in the topological regime to a simplified Hamiltonian $\Htoy$ in which the (quasi-)continuum of the wires for energies $|E| > \Delta$ is projected out. As such, $\Htoy$ includes one Majorana for each wire, $\gamL$  and $\gamR$, and a single fermionic orbital $f^\dagger_\sigma,f_\sigma$ with majority/minority spin $\sigma = \pm$, corresponding to the direction anti-parallel($+$) and parallel($-$) to the wire. For simplicity, we assume the Majoranas to be polarized anti-parallel to the external field in this simple model. The dot is subject to strong onsite Coulomb interaction $U/\Delta \gg 1$ and couples via spin-conserving and spin-flip tunneling to both Majoranas. 
Introducing $n_\sigma = d^\dagger_\sigma d_\sigma$ and the nonlocally fused fermion $\fMdag = (\gamL - i\gamR)/\sqrt{2}$ with $\nM = \fMdag\fM$, the 3-body Hamiltonian reads
\begin{align}
 \Htoy &= \sum_{\sigma=\pm} (\epsilon - \sigma B\sinfn{\Theta})n_\sigma + Un_+n_-\notag\\
 &+ \sinfn{\frac{\delphi}{4}}\sqbrack{\fMdag \nbrack{\ttoy f_+ + \Jtoy f^\dagger_-} + \Hc}\notag\\
 &+ \cosfn{\frac{\delphi}{4}}\sqbrack{i\fMdag \nbrack{\ttoy f^\dagger_+ - \Jtoy f_-} + \Hc}.\label{eq_toy}
\end{align}
The fused fermion has no ``onsite''-energy term since the Majoranas on their own are zero-energy modes, and since direct hybridization between the two Majoranas --- locally separated by the dot --- is negligible compared to the coupling between Majoranas and dot.

%%%%%%%%%%%%%%%%%%%%%%%%%%%%%%%%%%%%%%%%%%%%%%%%%%%%%%%%%%%%%
\begin{figure*}
	\centering
	\includegraphics[width=\linewidth]{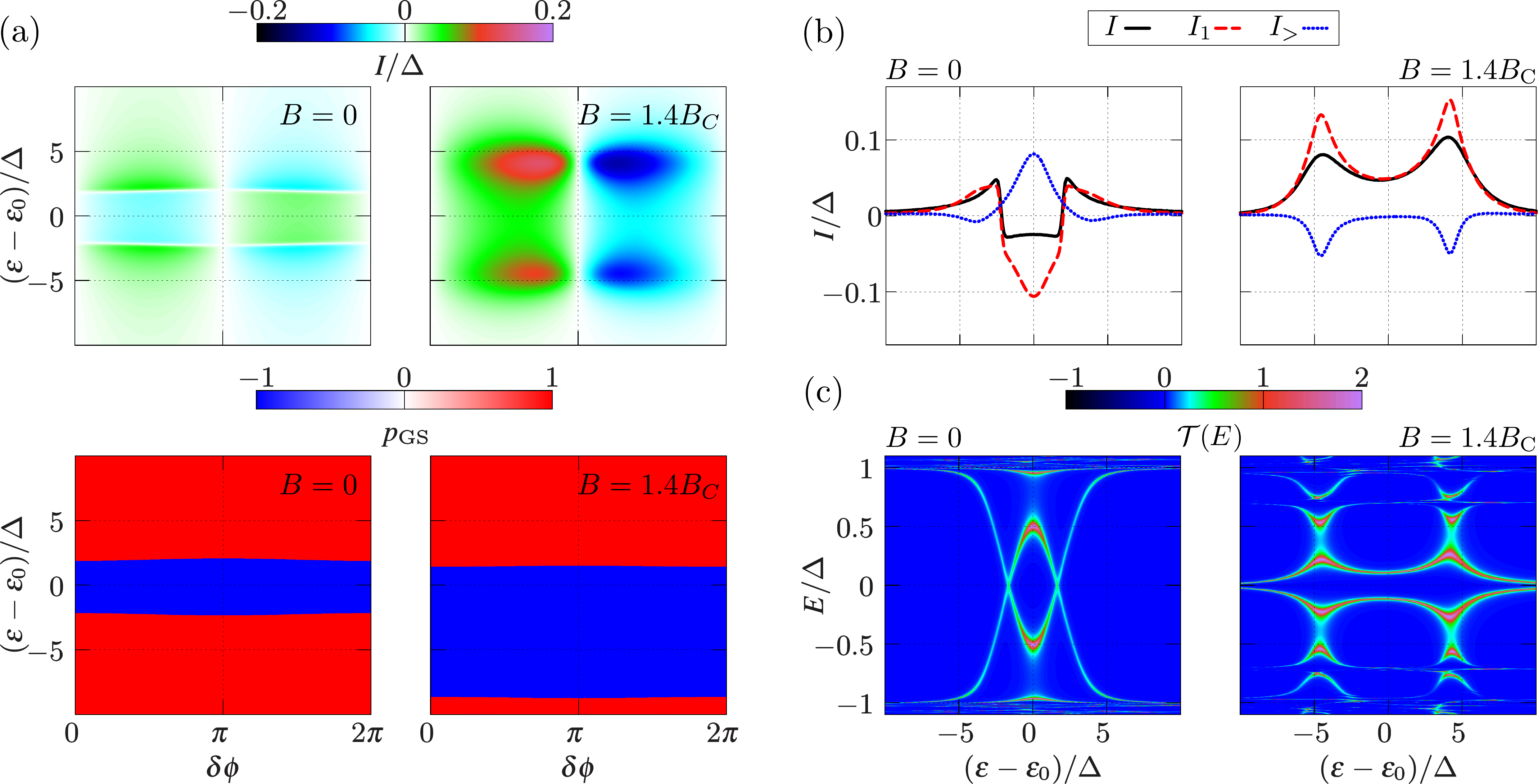}
	\caption{(a) $\delphi,\epsilon$-dependence of equilibrium supercurrent $I$ \BrackEq{eq_current} (upper panels) and ground state fermion parity $\pGS$ of simulated Hamiltonian \eq{eq_model} (lower panels) in the trivial $(B = 0)$ and topological $(B = 1.4B_C)$ regime.
	(b) Contribution from lowest subgap state $I_{1}$ (excluding outer Majoranas) and all higher levels $I_{>}$ to $I$ \BrackEq{eq_current_general} for $\delphi = \pi/2$. (c) Quasi-particle spectra exhibited by the transmission coefficient $\mathcal{T}$ of normal lead as obtained from \eq{eq_scattering} as a function of $\epsilon$ and transmission energy $E$ ($\mu \equiv 0$) for $\delphi = \pi/2$.
	Parameters are $M = 2000$, $l = 250\unit{nm}$, $\lW = 2700\unit{nm}$, $\lB = 50\unit{nm}$,$\delta B = 3\Delta$, $\delta \VB = 10\Delta$, $\epsilon_0 = -269.1\Delta$, $\meff = 0.026m_e$ with electron mass $m_e$, $\alpha = 16\unit{meV\,nm}$, $\Delta = 0.2\unit{meV}$, $\mu = 0.4\unit{meV}$, $T = 0.1\text{K} \approx 0.043\Delta$. We use $100$ positive and $100$ corresponding negative Matsubara frequencies to evaluate \Eq{eq_current}. For (a,b), we set $t_{\text{n}} = J_{\text{n}} = 0$; for (c), we set $t_{\text{n}}\sqrt{\Delta\rho} = 0.001t$, $J_{\text{n}}\sqrt{\Delta\rho} = 0.001 J$.
	}
	\label{fig_simulation}
\end{figure*}
%%%%%%%%%%%%%%%%%%%%%%%%%%%%%%%%%%%%%%%%%%%%%%%%%%%%%%%%%%%%%

The low-energy projection of the equilibrium supercurrent \eq{eq_current} is given by
\begin{equation}
 \Itoy = \Re\sqbrack{\exp\nbrack{i\frac{\delphi}{4}}\eqavg{\nbrack{\ttoy f^\dagger_+ - \Jtoy f^\dagger_-}\nbrack{\fMdag + \fM}}}.\label{eq_current_toy}
\end{equation}
To qualitatively compare this to the current $I$ in the simulation \BrackEq{eq_current}, $\dtoy$ and $\alphaToy$ entering the couplings $\ttoy = 1/(2\meff\dtoy^2), \Jtoy = \alphaToy/(2\dtoy)$ are set such that the peaks in $\Itoy$ as a function of $\delta\phi$ and $\epsilon$ for fixed Zeeman energy $\Btoy$ and $U$ deviate $\lesssim 1\%$ in height from those of $I$.  
Furthermore, for a qualitative assessment of the mean-field treatment of the interaction in the simulations, we evaluate the grandcanonical ensemble averages $\langle\dotsc\rangle_{\text{eq}} \sim e^{-\Htoy/T}$ with the two-particle dot onsite interaction $\sim U > 0$ fully accounted for, exploiting that $\Htoy$ decouples into two $4\times4$-blocks of opposite fermion parity. Note, however, that $\Htoy$ does not capture the internal spin-orbit coupling affecting hopping \emph{within} the normal region in the simulation. This leads to an underestimated magnitude of minority-spin tunneling and to a smaller level splitting compared to the simulation when choosing $\delta B = U/2$ and $\Btoy = B$. We compensate this by enhancing $\Jtoy$ with a factor $\alphaToy > \alpha$, and by setting $\delta B < U/2$.

\section{Results}
\label{sec_results}

\subsection{Supercurrent and parity flips}
\label{sec_parity_flip}
Figure~\ref{fig_simulation}(a) compares the equilibrium supercurrent $I(\delphi,\epsilon)$ obtained from the simulation in the topological regime $B > B_C$ to the one in the trivial regime $B < B_C$. Most noticeable in the latter are sharp steps between positive and negative currents as a function of $\epsilon$ for constant $\delphi$. These well-understood $\pi$-phase shifts of the supercurrent~\cite{Spivak1991Feb,Rozhkov2001Nov,vanDam2006Aug} directly coincide with flips of the fermion-parity in the junction, reflecting also in the total ground state parity $\pGS$. Given a splitting $\sim 2(B + \delta B)$ due to both magnetic field and interaction, the resonances of the two levels with the chemical potential define an $\epsilon$-interval in which this parity is odd, and the current sign is inverted. 
By striking contrast, in the topological regime $B > B_C$, there is no sign change as a function of $\epsilon$, and instead we find two peaks and a more abrupt, non-sinusoidal sign change at phase $\delphi = \pi$. The ground state parity $\pGS$ still flips at two different levels $\epsilon$, but not anymore close to the current peak positions.

The question when and why flips of which parities correspond to a rapid sign change in the supercurrent can be answered with the general current expression
\begin{equation}
 I = 2\left\langle\frac{\partial H}{\partial\delphi}\right\rangle_{\text{eq}} \rightarrow -\sum_{\epsilon_s \geq 0}\tanhfn{\frac{\epsilon_s}{2T}}\frac{\partial\epsilon_s}{\partial\delphi}.\label{eq_current_general}
\end{equation}
that holds within the mean-field treatment of the Coulomb interaction in the dot adopted here. The $\epsilon_s$ are the quasi-particle energies with respect to the chemical potential $\mu$, i.e., the eigenvalues of the Bogoluibov-de Gennes Hamiltonian $H$. These always come in pairs $\epsilon_{s,\pm}$ with opposite signs, $\epsilon_{s,+} = -\epsilon_{s,-}$ representing particles and holes with opposite phase derivatives, $\partial\epsilon_{s,+}/\partial\delphi = -\partial\epsilon_{s,-}/\partial\delphi$. Given a grand-canonical ensemble at low temperatures, the parity of a single state $s$ flips if its energies $\epsilon_{s,\pm}$ as a function of the system parameters cross zero and change sign. 
Since \Eq{eq_current_general} sums only over non-negative energies, the contribution of the state $s$ then swaps between the one from $\epsilon_{s,+}$ and from $\epsilon_{s,-}$, thereby leading to a sign change due to the opposite sign of $\partial\epsilon_{s}/\partial\delphi$. This in any case results in a (temperature broadened) step in the supercurrent. If the particular state $s$ gives the dominant contribution to $I$, meaning it has a relatively large phase derivative, this step-like transition even causes an overall sign change of $I$.

The main point is now that the energies with the largest phase derivative belong to the subgap states localized close to the dot-wire interfaces. In \Fig{fig_simulation}(c), we plot the $\epsilon$-dependence of these subgap-state energies as probed by the transmission $\mathcal{T}(E)$ obtained from \Eq{eq_scattering}. In the trivial regime $B = 0$, there are two Yu-Shiba-Rusinov subgap states~\cite{Yu1965,Shiba1968Sep,Rusinov1969} forming in a dot with effective charging energy larger than the gap, $2\delta B = 6\Delta \gg \Delta$. These states cross zero at the two levels $\epsilon$ at which \Fig{fig_simulation}(a) indicates parity flips with concomitant supercurrent sign changes. In the topological regime $B > B_{\text{C}}$, the visible energies in \Fig{fig_simulation}(c) do not cross zero anymore, and instead form a diamond shape.
This shape has previously been highlighted in the context of how Majorana fermions and their nonlocality influence the quasi-particle spectra~\cite{Prada2017Aug,Deng2018Aug}. Here, \Fig{fig_simulation}(b,c) show that these states provide the dominant current contribution and do not flip parity as a function of $\epsilon$, thereby also leading to an absent sign change in the supercurrent.

The abrupt current sign flip at $\delphi = \pi$ has previously been addressed in, e.g., \Ref{Cayao2018May}. It stems from the fact that the Majoranas in the wire decouple from the dot at phase difference $\delphi = \pi$, typically causing a zero-energy crossing and thus a parity flip. As we see in the next section \ref{sec_toy}, this parity flip is localized to the Majoranas close to the dot. Moreover, as the constant ground-state parity $\pGS$ around $\delphi = \pi$ suggests, it is compensated by another parity flip of the Majoranas at the outer ends of the wires, which for any \emph{finite} wire length have a strongly suppressed, yet not exactly vanishing phase dependence.

As further illustrated in the next section \ref{sec_toy}, the absence of step-like $\epsilon$-dependences (leading to current sign changes) in the topological phase can be explained by the effect of hybridization and level repulsion. In the trivial regime, the wire states are sufficiently far away from $E = 0$ on the scale of the hybridization with the dot. The dot levels increasing linearly with $\epsilon$ around $E = 0$ can therefore cross $0$ without being level-repelled, and this crossing leads to a parity flip and a current sign change in the grandcanonical equilibrium. In the topological regime, two of the altogether 4 Majorana modes are, however, close to the dot, and therefore hybridize enough with the dot to repel the dot levels from zero energy.
The total ground-state parity $\pGS$ still changes, but only due to the parity change in the weakly phase-dependent states formed by the Majoranas at the outer ends of the wires, which, as stated above, are practically irrelevant for the supercurrent by \Eq{eq_current_general}. This also explains why, unlike for the trivial regime, the level positions at which the $\pGS$-flips in the lower-right panel of \Fig{fig_simulation}(a) are located do not coincide with the current peaks in the topological regime.

Our main finding is thus that Majorana modes hybridizing with the quantum dot formed by the normal junction prevents parity flips in and around the dot through level repulsion, thereby avoiding abrupt supercurrent sign changes as a function of the dot level. We, however, stress and show in the following section~\ref{sec_toy} that this is only related to the existence of wire states close to zero energy on the scale of their coupling to the dot states. The logic is thus that while the \emph{presence} of an \emph{abrupt} sign change as a function $\epsilon$ indicates the non-existence of robust zero energy states at the dot-wire boundary, an \emph{absence of} a sign change or any step-like transition with or without overall sign change does not imply the existence of Majoranas.  

\subsection{Low-energy approximation}
\label{sec_toy}

%%%%%%%%%%%%%%%%%%%%%%%%%%%%%%%%%%%%%%%%%%%%%%%%%%%%%%%%%%%%%
\begin{figure}
	\centering
	\includegraphics[width=\linewidth]{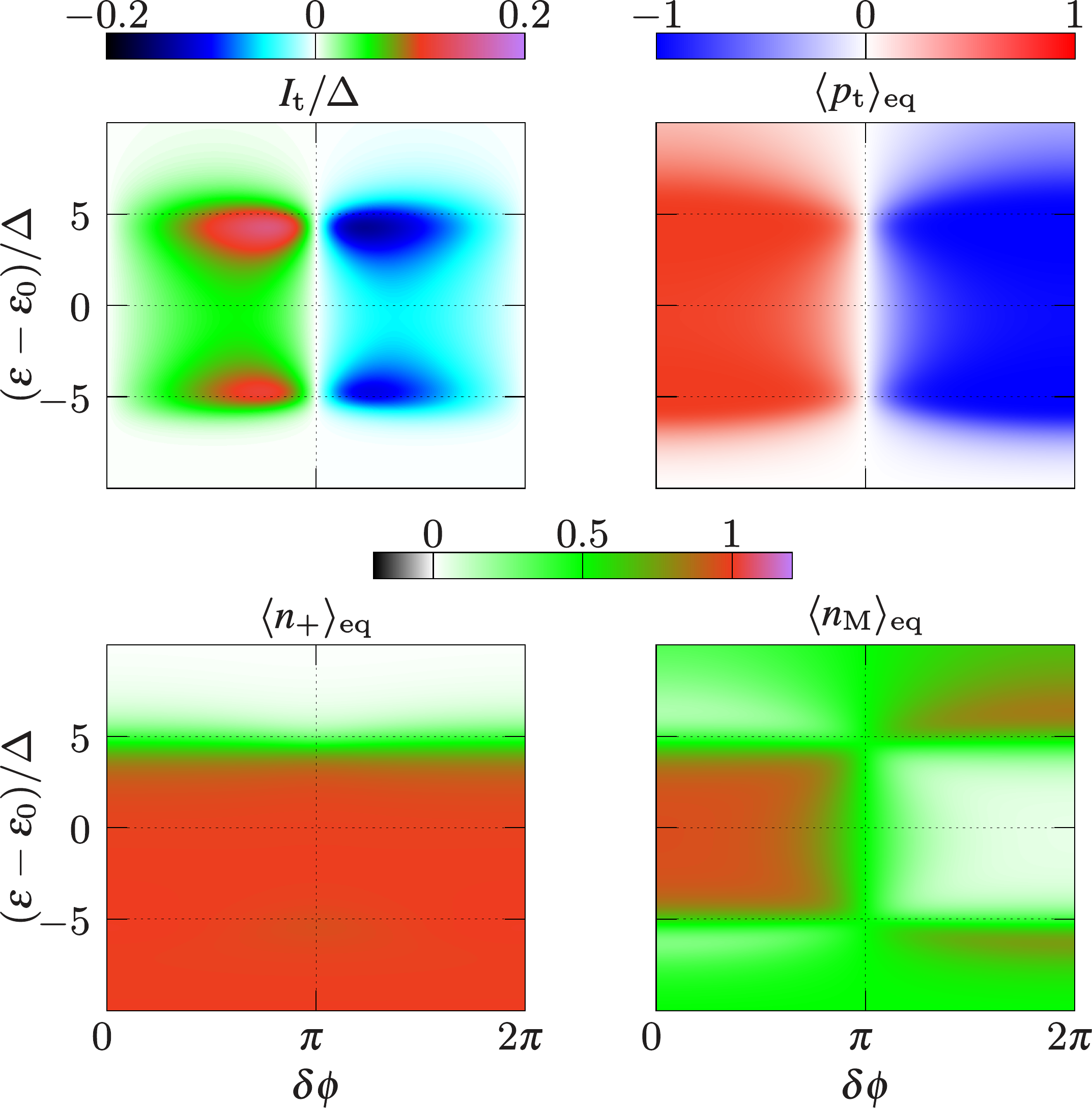}
	\caption{Equilibrium supercurrent $\Itoy$, equilibrium parity $\langle\ptoy\rangle_{\text{eq}}$, and equilibrium occupations $\langle n_+\rangle_{\text{eq}}$, $\langle\nM\rangle_{\text{eq}}$ of the simplified model \eq{eq_toy} as a function of phase difference $\delphi$ and dot level $\epsilon$. Parameters are set to $\dtoy = 102.5\unit{nm}$, $\alphaToy = 1.6\alpha$, $U = 8.6\Delta$, $\epsilon_0 = -4\Delta$, $\meff = 0.026m_e$, $\alpha = 16\unit{meV\,nm}$, $\Delta = 0.2\unit{meV}$, $\mu = 0.4\unit{meV}$, $T = 0.1\text{K} \approx 0.043\Delta$.
	}
	\label{fig_toy}
\end{figure}
%%%%%%%%%%%%%%%%%%%%%%%%%%%%%%%%%%%%%%%%%%%%%%%%%%%%%%%%%%%%%

We now turn to low-energy approximations to get a better physical picture of the current-carrying subgap states in both the trivial and topological regime. We start with the latter, topological case, which is captured by the Hamiltonian \eq{eq_toy}; important expectation values are plotted in \Fig{fig_toy}. 
Exhibiting good qualitative agreement between the low-energy supercurrent $\Itoy(\epsilon,\delphi)$ and the current $I(\epsilon,\delphi)$ obtained from the simulation, the plots demonstrate in particular that it is energetically favorable for the majority-spin parity $p_+ = 1 - 2n_+$ to flip \emph{together with} the parity of the nonlocally fused fermion $\pM = 1 - 2\nM$ upon crossing $\epsilon_+ = \epsilon - B = 0$. For $-B - U < \epsilon \lesssim B$ and, thus, suppressed spin-flip tunneling, this is due to direct tunneling $\sim\ttoy$ dominating the Hamiltonian \eq{eq_toy} by pairing $\sim (\fMdag f^\dagger_+  + \text{H.c.})$ for phases $0 < \delphi < \pi$, and by regular exchange $\sim (\fMdag f_+  + \text{H.c.})$ for $\pi < \delphi < 2\pi$ with opposite $\pM$.
The situation close to the second resonance $\epsilon + B + U = 0$ depends on the ratio $\Jtoy/\ttoy$ quantifying the relevance of spin-flip tunneling. In case of comparable or large spin-flip contributions $\Jtoy/\ttoy \sim 1$ as chosen in \Fig{fig_simulation}, the terms $\sim \cosfn{\frac{\delphi}{4}}(f^\dagger_-\fM + \Hc)$ and, respectively, $\sim \sinfn{\frac{\delphi}{4}}(f_-\fM + \Hc)$ in the Hamiltonian \eq{eq_toy} cause $\pM$ to flip along with $p_-$ for an extended regime $\epsilon < -U - B$. 
Altogether, we thus conclude that the above pointed out lack of $\epsilon$-dependent parity flips in the eigenstates localized around the dot effectively results from a parity flip in \emph{both} the bare dot state and in the fused fermion.
Furthermore, we see explicitly that the sharp sign change in $I$ at phase $\delphi = \pi$ observed in the simulation \BrackFig{fig_simulation} is caused by the parity $\pM$ of the nonlocally fused Majoranas flipping~\cite{Cayao2018May}. This parity $\pM$ in fact has a $4\pi$-periodicity in $\delphi$ which, however, is not observable in the total ground-state parity $\pGS$ because the parity of the Majoranas at the outer ends of the wires flips along with $\pM$.

Finally, to illustrate the level-repulsion effect on the supercurrent for increasing magnetic fields, we consider a toy model described by the Hamiltonian \eq{eq_model} with only $1$ site per wire and dot $(M = 3)$, and with lattice spacing $d = \dtoy$ as well as spin-orbit coupling constant $\alpha = \alphaToy$ chosen equal to those for the low-energy model in the topological regime. This approximates the case in which the transport \emph{in the wires} is dominated by regular Andreev subgap wire states. Namely, while zero-energy modes exist at $B$ equal to $B_0 = \sqrt{\Delta^2 + (2t - \mu)^2}$, these states only correspond to Majorana excitations at the singular parameter point $2t = \mu$, which is avoided in the following.

%%%%%%%%%%%%%%%%%%%%%%%%%%%%%%%%%%%%%%%%%%%%%%%%%%%%%%%%%%%%%
\begin{figure}
	\centering
	\includegraphics[width=\linewidth]{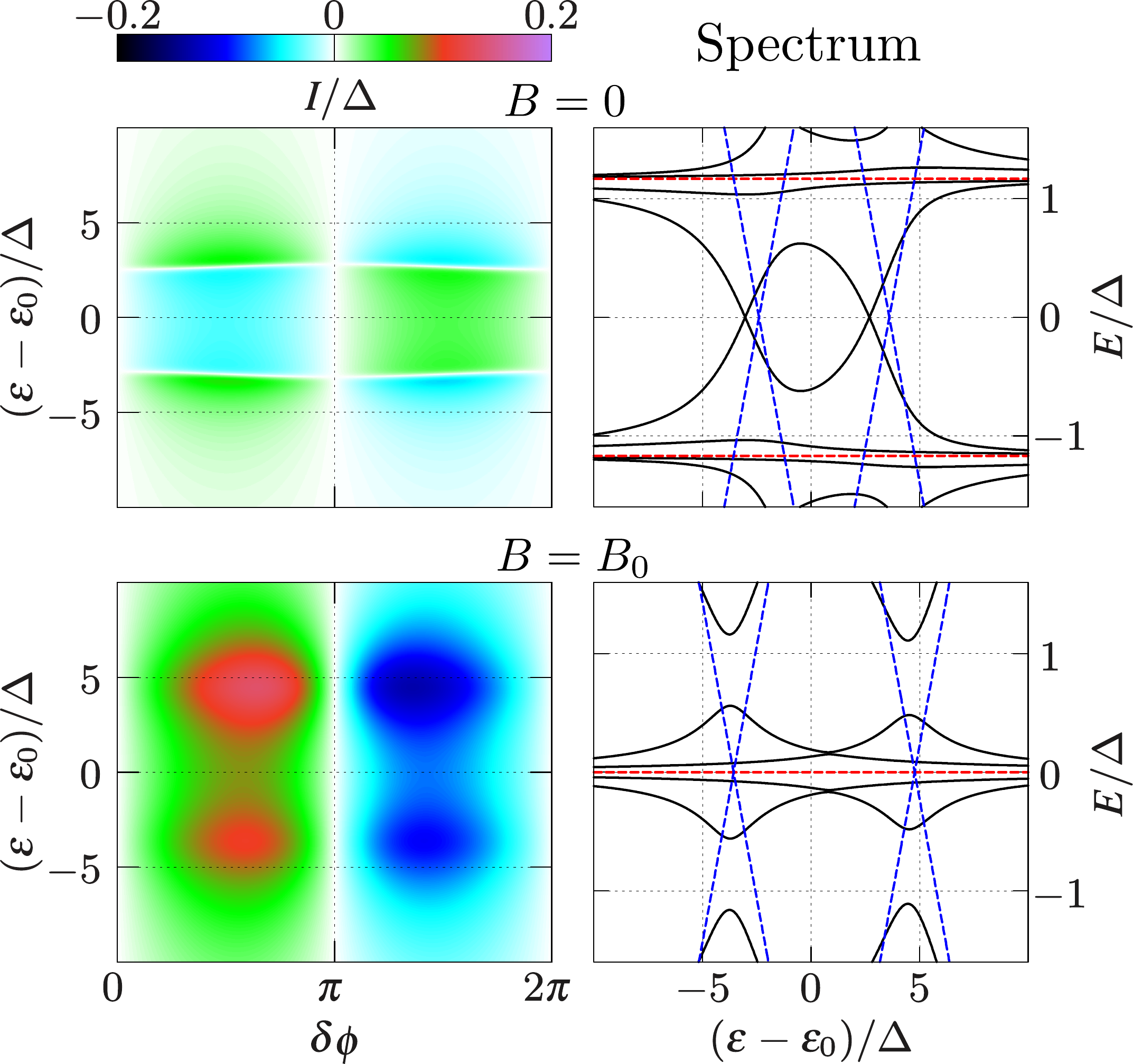}
	\caption{Equilibrium supercurrent $I$ and quasi-particle spectrum as a function of dot level $\epsilon$ and/or phase difference $\delphi$ of the system \eq{eq_model} with only site per wire and one site for dot ($M = 3$). The spectra are plotted for a phase difference $\delphi = \pi/2$. The blue/red lines show the bare energies of the dot/wire when uncoupled from each other. Parameters apart from $B$ are taken from \Fig{fig_simulation}(a), but with $d \rightarrow \dtoy = 102.5\unit{nm}$, $\alpha \rightarrow \alphaToy = 1.6\alpha$, and $\epsilon_0 = 0$.
	}
	\label{fig_non_topo}
\end{figure}

In \Fig{fig_non_topo}, we plot both the supercurrent $I$ and the spectrum of the Bogoluibov-de Gennes Hamiltonian $H$ as function of $\epsilon$ and $\delphi$ for $B = 0$ and $B = B_0$ with $2\ttoy - \mu \approx -0.6\Delta \neq 0$. Without magnetic field, we find --- similarly to the case $M = 2000$ in \Fig{fig_simulation} -- an abrupt sign change at level positions $\epsilon$ corresponding to zero crossings in the spectrum. For $B = B_0$, these crossings do not appear anymore, since each wire features a \emph{non-Majorana} state at zero energy (red lines) that level-repels the bare dot levels (blue lines). Consequently, no abrupt step and hence sign change is found in the supercurrent anymore.

\section{Summary and outlook}

This paper has investigated how the stationary supercurrent through two superconducting wires coupled by a quantum dot is affected by the presence of Majorana zero modes in the wires. We have found that such modes prohibit step-like features including current sign reversals upon parity changes in the dot --- the formation of so-called $\pi$-junctions --- that are well-known to occur in the absence of zero-energy wire states. The disappearance of such steps in the current can, however, be explained entirely by the fact that zero-energy excitations prevent parity flips through level-repulsion; whether or not the zero-energy state is a Majorana does not seem to be essential. We thus conclude that the absence of sign reversals in the supercurrent does not always enable a clear identification of Majorana modes, but it allows to rule out their existence, and thereby still provides a useful test criterion for experiments.

\acknowledgments
The research was supported by the Danish National Research Foundation, the Danish Council for Independent Research $\vert$ Natural Sciences, and the Microsoft Corporation.
We acknowledge fruitful discussions with A. Whiticar, A. Fornieri, D.  Razmadze, E. O'Ferrell, C. Marcus and useful comments from J. Cayao.

\section*{Appendix}

\appendix

\section{Equilibrium averages, Majorana basis and matrix inversions}
\label{app_eq}
Let us sketch how to evaluate the equilibrium averages $\langle\dotsc\rangle_{\text{eq}}$ entering the current formula \eq{eq_current}. We employ a method analogous to appendix A of \Ref{Novotny2005Dec}, applying it to our finite 1D wire-junction-wire system. It starts by discretizing and second-quantizing the Hamiltonian \eq{eq_model},
\begin{align}
 \hat{H} &= \sum_{j=1}^{M}\clbrack{\sum_{\sigma=\uparrow,\downarrow}\sqbrack{(2t - \mu + V_j)c^\dagger_{j\sigma}c_{j\sigma}}}\notag\\
 &\phantom{=}\crbrack{\phantom{\sum_i}+ \sqbrack{B_j c^\dagger_{j\uparrow}c_{j\downarrow} + \Delta_je^{i\phi_j}c^\dagger_{j\uparrow}c^\dagger_{j\downarrow} + \text{H.c.}}}\notag\\
 &+\sum_{j=1}^{M-1}\clbrack{\sqlbrack{\nbrack{-t\sum_{\sigma=\uparrow,\downarrow}c^\dagger_{j\sigma}c_{j+1\sigma}}}} \notag\\
 &\phantom{+}\crbrack{\phantom{\clbrack{\sum_{j=1}^{M-1}}} + \sqrbrack{J\nbrack{c^\dagger_{j\uparrow}c_{j+1\downarrow} - c^\dagger_{j\downarrow}c_{j+1\uparrow}}} + \text{H.c.}},\label{eq_hamiltonian_quantized}
\end{align}
introducing creation and annihilation operators $c^\dagger_{j\sigma}$ for spin $\sigma=\uparrow,\downarrow$ at site $j = 1,\dotsc,M$, labeling discrete points $x_j$ from left to right in our system. The symbols $V_j = V(x_j)$, $B_j = B(x_j)$, $\Delta_j = \Delta(x_j)$, $\phi_j = \phi(x_j)$ denote, respectively, the corresponding potential, Zeeman field, superconducting gap and superconducting phase at these points, as given in \Fig{fig_setup} and in \Sec{sec_model}. The sets of points $x_j$ belonging to the left wire, dot, and right wire are determined by the total number of sites $M$ weighted by the corresponding length ratios.\par

For more efficient numerical computations, we switch to the Majorana basis by introducing
\begin{equation}
 \gamma_{j\eta\sigma} = \frac{\delta_{\eta+} - i\delta_{\eta-}}{\sqrt{2}}\nbrack{e^{-i\frac{\phi_j}{2}}c_{j\sigma} + \eta e^{i\frac{\phi_j}{2}}c^\dagger_{j\sigma}}.\label{eq_majoranas}
\end{equation}
The operators \eq{eq_majoranas} are hermitian, $\gamma_{j\eta\sigma}^\dagger =  \gamma_{j\eta\sigma}$ and fulfill the anti-commutation relations $\{\gamma_{j\eta\sigma},\gamma_{j'\eta'\sigma'}\} = \delta_{jj'}\delta_{\eta\eta'}\delta_{\sigma\sigma'}$. The backtransform to regular creation and annihilation operators reads
\begin{align}
 c_{j\sigma} &= \frac{1}{\sqrt{2}}\nbrack{\gamma_{j+\sigma} + i\gamma_{j-\sigma}}e^{i\frac{\phi_j}{2}}.\label{eq_backtransform}
\end{align}

We define the imaginary-time, Matsubara Green's function for the Majoranas \eq{eq_backtransform} as
\begin{equation}
 G_{j\eta\sigma,j'\eta'\sigma'}(\tau,\tau') = -\Tr\cbrack{\mathcal{T}_{\tau}\sqbrack{\gamma_{j\eta\sigma}(\tau)\gamma_{j'\eta'\sigma'}(\tau')}\rho_{\text{eq}}}\label{eq_greens_function},
\end{equation}
where $\rho_{\text{eq}} = e^{-\hat{H}/T}/\Tr\sqbrack{e^{-\hat{H}/T}}$ is the equilibrium state, $\mathcal{T}_\tau$ is the time-ordering operator that shifts operators at larger $\tau$ further to the left, and $\gamma_{j\eta\sigma}(\tau) = e^{\hat{H}\tau}\gamma_{j\eta\sigma}e^{-\hat{H}\tau}$. 
The Majorana equilibrium average is thus 
\begin{equation}
 \left\langle\gamma_{j\eta\sigma}\gamma_{j'\eta'\sigma'}\right\rangle_{\text{eq}} = -G^0_{j\eta\sigma,j'\eta'\sigma'} := -\lim_{\tau\rightarrow 0_+}G_{j\eta\sigma,j'\eta'\sigma'}(\tau,0).
\end{equation}
The equilibrium averages entering the current \eq{eq_current} then follow from \Eq{eq_backtransform}:
\begin{align}
 \langle c^\dagger_{j\sigma} c_{j'\sigma'}\rangle_{\text{eq}} &= -\frac{1}{2}\sqlbrack{G^0_{j+\sigma,j'+\sigma'} + G^0_{j-\sigma,j'-\sigma'}}\label{eq_equi_average_temp}\\
 &\phantom{-\frac{1}{2}\sqlbrack{}}+\sqrbrack{iG^0_{j+\sigma,j'-\sigma'} - iG^0_{j-\sigma,j'+\sigma'}}e^{i\frac{\phi_j - \phi_{j'}}{2}}.\notag
\end{align}
Since the Hamiltonian \eq{eq_hamiltonian_quantized} is quadratic in the fields, the full matrix of Green's functions $G(\tau,0)$ and hence $G^0$ (i.e., $G^0_{j\eta\sigma,j'\eta'\sigma'} = (G^0)_{j\eta\sigma,j'\eta'\sigma'}$) can be obtained from the equation of motion
\begin{align}
 \frac{d}{d\tau}&G_{j\eta\sigma,j'\eta'\sigma'}(\tau,0) = -\delta(\tau)\delta_{jj'}\delta_{\eta\eta'}\delta_{\sigma\sigma'}\notag\\
 &- \sum_{\tilde{j}=1}^{M}\sum_{\tilde{\eta}=\pm}\sum_{\tilde{\sigma}=\uparrow,\downarrow}\mathcal{H}_{j\eta\sigma,\tilde{j}\tilde{\eta}\tilde{\sigma}}\cdot G_{\tilde{j}\tilde{\eta}\tilde{\sigma},j'\eta'\sigma'}(\tau,0),\label{eq_of_motion}
\end{align}
with the matrix $\mathcal{H}$ identified from the commutator of the fields with the Hamiltonian, which has the general form
\begin{equation}
 [\gamma_{j\eta\sigma},\hat{H}] = \sum_{\tilde{j}=1}^{M}\sum_{\tilde{\eta}=\pm}\sum_{\tilde{\sigma}=\uparrow,\downarrow}\mathcal{H}_{j\eta\sigma,\tilde{j}\tilde{\eta}\tilde{\sigma}}\cdot \gamma_{\tilde{j}\tilde{\eta}\tilde{\sigma}}.
\end{equation}
Equation \eq{eq_of_motion} is solved by first solving the corresponding algebraic equation in the Matsubara frequency domain $(d/d\tau \rightarrow -i\omega_n)$ and then by backtransforming to the matrix of Green's functions $G^0$ at $\tau \rightarrow 0_+$:
\begin{equation}
 G^0 = T\sum_{n=-\infty}^\infty\frac{1}{i\omega_n - \mathcal{H}} \quad,\quad \omega_n = T\pi(2n + 1)\,\,,\,\, n\in\mathds{Z}.\label{eq_green_matrix_temp}
\end{equation}
Numerical simplifications now arise from the hermiticity of the Majoranas \eq{eq_majoranas}, leading to $\mathcal{H} = i\tilde{\mathcal{H}}$ with $\Im\tilde{\mathcal{H}} = 0$ and $\tilde{\mathcal{H}}^T = -\mathcal{H}$ ($\bullet^T = $ transposition). Since inversion and transposition commute, and since every positive frequency in the sum in \Eq{eq_green_matrix_temp} has a corresponding negative frequency with equal absolute value, we find
\begin{equation}
 G^0 = -iT\tilde{G}^0 \quad,\quad \tilde{G}^0 = \sum_{n=0}^\infty\sqbrack{\frac{1}{\omega_n - \tilde{\mathcal{H}}} - \nbrack{\frac{1}{\omega_n - \tilde{\mathcal{H}}}}^T}.\label{eq_green_matrix}
\end{equation}
The matrix inversion in \Eq{eq_green_matrix} is performed only once for positive frequencies, and only on a purely real matrix, which is numerically faster than the complex matrix inversion in \Eq{eq_green_matrix_temp} for both positive and negative frequencies. Moreover, since $\hat{H}$ only couples nearest neighbor sites, $(\omega_n - \tilde{\mathcal{H}})$ has a simple block-band structure,
\begin{align}
&\mathcal{R} = \omega_n - \tilde{\mathcal{H}} =\label{eq_block_band}\\
 &\begin{pmatrix}  &  &  &  &  & \vdots & \vdots & \vdots & \vdots & \vdots \\ \cdots & 0 & \ddots & \ddots & \ddots & 0 & 0 & 0 & 0 & \cdots \\ \cdots & 0 & 0 & \mathcal{R}^T_{c,j-1} & \mathcal{R}_{d,j} & -\mathcal{R}_{c,j} & 0 & 0 & 0 & \cdots \\ \cdots & 0 & 0 & 0 & \mathcal{R}^T_{c,j} & \mathcal{R}_{d,j+1} & -\mathcal{R}_{c,j+1} & 0 & 0 & \cdots\\ \cdots & 0 & 0 & 0  &  0 & \ddots & \ddots & \ddots & 0 & \cdots \\ \vdots & \vdots & \vdots & \vdots & \vdots &  &  &  &  & \end{pmatrix}\notag
\end{align}
with the $4\times 4$ subblocks
\begin{align}
 &\mathcal{R}_{d,j}\\
 &= \begin{pmatrix}\omega_n & \mu - 2t - V_j & 0 & \Delta_j - B_j \\ 2t - \mu + V_j & \omega_n & \Delta_j + B_j & 0 \\ 0 & -\Delta_j - B_j & \omega_n & \mu - 2t - V_j \\ B_j - \Delta_j & 0 & 2t - \mu + V_j & \omega_n\end{pmatrix}\notag
\end{align}
and
\begin{align}
 \mathcal{R}_{c,j} &= \begin{pmatrix}t_{j-} & -t_{j+} & -J_{j-} & J_{j+} \\ t_{j+} & t_{j-} & -J_{j+} & -J_{j-} \\ J_{j-} & -J_{j+} & t_{j-} & -t_{j+} \\ J_{j+} &  J_{j-} & t_{j+} & t_{j-}\end{pmatrix}
\end{align}
in the Majorana basis $(+\uparrow,-\uparrow,+\downarrow,-\downarrow)$, where 
\begin{align}
t_{j\pm} = t\cosfn{\frac{\phi_j - \phi_{j+1}}{2} + \frac{\pm \pi - \pi}{4}}\notag\\
J_{j\pm} = J\cosfn{\frac{\phi_j - \phi_{j+1}}{2} + \frac{\pm \pi - \pi}{4}}.
\end{align}
Due to this block-band structure \eq{eq_block_band}, the $4\times 4$ subblocks of the full inverses in $\tilde{G}_0 = \sum_n(\omega_n - \tilde{\mathcal{H}})^{-1}$ required to evaluate the averages in the current formula \eq{eq_current} using \Eq{eq_equi_average_temp} and \Eq{eq_green_matrix} can be efficiently determined by iteratively applying the block-inversion scheme  
\begin{align}
 &\begin{pmatrix}A & B \\ C & D\end{pmatrix}^{-1} = \label{eq_block_inversion}\\
 &\begin{pmatrix}(A - BD^{-1}C)^{-1} & -(A - BD^{-1}C)^{-1}BD^{-1} \\  -(D - CA^{-1}B)^{-1}CA^{-1} & (D - CA^{-1}B)^{-1}\end{pmatrix}\notag
\end{align}
with matrices $A,B,C,D$. Finally, we also use the Majorana basis \eq{eq_majoranas} and the block inversion \eq{eq_block_inversion} for
\begin{equation}
 \frac{1}{E - \mathcal{H} + i\pi WW^\dagger} \overset{\mathcal{H} = i\tilde{\mathcal{H}}}{=} \frac{i}{iE + \tilde{\mathcal{H}} - \pi WW^\dagger},\label{eq_resolvent}
\end{equation}
in \Eq{eq_scattering}. Due to the broadening $i\pi WW^\dagger$ from the normal lead, the matrix to invert is not real-valued, but the result for $-E$ is simply obtained from \Eq{eq_resolvent} for $+E$ by complex conjugation, as $\Im(WW^\dagger) = 0$ both in the particle-hole \BrackEq{eq_coupling} and Majorana basis. 


\begin{thebibliography}{93}%
\makeatletter
\providecommand \@ifxundefined [1]{%
 \@ifx{#1\undefined}
}%
\providecommand \@ifnum [1]{%
 \ifnum #1\expandafter \@firstoftwo
 \else \expandafter \@secondoftwo
 \fi
}%
\providecommand \@ifx [1]{%
 \ifx #1\expandafter \@firstoftwo
 \else \expandafter \@secondoftwo
 \fi
}%
\providecommand \natexlab [1]{#1}%
\providecommand \enquote  [1]{``#1''}%
\providecommand \bibnamefont  [1]{#1}%
\providecommand \bibfnamefont [1]{#1}%
\providecommand \citenamefont [1]{#1}%
\providecommand \href@noop [0]{\@secondoftwo}%
\providecommand \href [0]{\begingroup \@sanitize@url \@href}%
\providecommand \@href[1]{\@@startlink{#1}\@@href}%
\providecommand \@@href[1]{\endgroup#1\@@endlink}%
\providecommand \@sanitize@url [0]{\catcode `\\12\catcode `\$12\catcode
  `\&12\catcode `\#12\catcode `\^12\catcode `\_12\catcode `\%12\relax}%
\providecommand \@@startlink[1]{}%
\providecommand \@@endlink[0]{}%
\providecommand \url  [0]{\begingroup\@sanitize@url \@url }%
\providecommand \@url [1]{\endgroup\@href {#1}{\urlprefix }}%
\providecommand \urlprefix  [0]{URL }%
\providecommand \Eprint [0]{\href }%
\providecommand \doibase [0]{http://dx.doi.org/}%
\providecommand \selectlanguage [0]{\@gobble}%
\providecommand \bibinfo  [0]{\@secondoftwo}%
\providecommand \bibfield  [0]{\@secondoftwo}%
\providecommand \translation [1]{[#1]}%
\providecommand \BibitemOpen [0]{}%
\providecommand \bibitemStop [0]{}%
\providecommand \bibitemNoStop [0]{.\EOS\space}%
\providecommand \EOS [0]{\spacefactor3000\relax}%
\providecommand \BibitemShut  [1]{\csname bibitem#1\endcsname}%
\let\auto@bib@innerbib\@empty
%</preamble>
\bibitem [{\citenamefont {Nayak}\ \emph {et~al.}(2008)\citenamefont {Nayak},
  \citenamefont {Simon}, \citenamefont {Stern}, \citenamefont {Freedman},\ and\
  \citenamefont {Das~Sarma}}]{Nayak2008Sep}%
  \BibitemOpen
  \bibfield  {author} {\bibinfo {author} {\bibfnamefont {C.}~\bibnamefont
  {Nayak}}, \bibinfo {author} {\bibfnamefont {S.~H.}\ \bibnamefont {Simon}},
  \bibinfo {author} {\bibfnamefont {A.}~\bibnamefont {Stern}}, \bibinfo
  {author} {\bibfnamefont {M.}~\bibnamefont {Freedman}}, \ and\ \bibinfo
  {author} {\bibfnamefont {S.}~\bibnamefont {Das~Sarma}},\ }\href {\doibase
  10.1103/RevModPhys.80.1083} {\bibfield  {journal} {\bibinfo  {journal} {Rev.
  Mod. Phys.}\ }\textbf {\bibinfo {volume} {80}},\ \bibinfo {pages} {1083}
  (\bibinfo {year} {2008})}\BibitemShut {NoStop}%
\bibitem [{\citenamefont {Wilczek}(2009)}]{Wilczek2009Sep}%
  \BibitemOpen
  \bibfield  {author} {\bibinfo {author} {\bibfnamefont {F.}~\bibnamefont
  {Wilczek}},\ }\href {\doibase 10.1038/nphys1380} {\bibfield  {journal}
  {\bibinfo  {journal} {Nat. Phys.}\ }\textbf {\bibinfo {volume} {5}},\
  \bibinfo {pages} {614} (\bibinfo {year} {2009})}\BibitemShut {NoStop}%
\bibitem [{\citenamefont {Franz}(2010)}]{Franz2010Mar}%
  \BibitemOpen
  \bibfield  {author} {\bibinfo {author} {\bibfnamefont {M.}~\bibnamefont
  {Franz}},\ }\href {\doibase 10.1103/Physics.3.24} {\bibfield  {journal}
  {\bibinfo  {journal} {Physics}\ }\textbf {\bibinfo {volume} {3}} (\bibinfo
  {year} {2010}),\ 10.1103/Physics.3.24}\BibitemShut {NoStop}%
\bibitem [{\citenamefont {Stern}(2010)}]{Stern2010Mar}%
  \BibitemOpen
  \bibfield  {author} {\bibinfo {author} {\bibfnamefont {A.}~\bibnamefont
  {Stern}},\ }\href {\doibase 10.1038/nature08915} {\bibfield  {journal}
  {\bibinfo  {journal} {Nature}\ }\textbf {\bibinfo {volume} {464}},\ \bibinfo
  {pages} {187} (\bibinfo {year} {2010})}\BibitemShut {NoStop}%
\bibitem [{\citenamefont {Leijnse}\ and\ \citenamefont
  {Flensberg}(2012)}]{Leijnse2012Nov}%
  \BibitemOpen
  \bibfield  {author} {\bibinfo {author} {\bibfnamefont {M.}~\bibnamefont
  {Leijnse}}\ and\ \bibinfo {author} {\bibfnamefont {K.}~\bibnamefont
  {Flensberg}},\ }\href {\doibase 10.1088/0268-1242/27/12/124003} {\bibfield
  {journal} {\bibinfo  {journal} {Semicond. Sci. Technol.}\ }\textbf {\bibinfo
  {volume} {27}},\ \bibinfo {pages} {124003} (\bibinfo {year}
  {2012})}\BibitemShut {NoStop}%
\bibitem [{\citenamefont {Kitaev}(2003)}]{Kitaev2003Jan}%
  \BibitemOpen
  \bibfield  {author} {\bibinfo {author} {\bibfnamefont {A.~{\relax Yu}.}\
  \bibnamefont {Kitaev}},\ }\href {\doibase 10.1016/S0003-4916(02)00018-0}
  {\bibfield  {journal} {\bibinfo  {journal} {Ann. Phys.}\ }\textbf {\bibinfo
  {volume} {303}},\ \bibinfo {pages} {2} (\bibinfo {year} {2003})}\BibitemShut
  {NoStop}%
\bibitem [{\citenamefont {Das~Sarma}\ \emph {et~al.}(2015)\citenamefont
  {Das~Sarma}, \citenamefont {Freedman},\ and\ \citenamefont
  {Nayak}}]{Sarma2015Oct}%
  \BibitemOpen
  \bibfield  {author} {\bibinfo {author} {\bibfnamefont {S.}~\bibnamefont
  {Das~Sarma}}, \bibinfo {author} {\bibfnamefont {M.}~\bibnamefont {Freedman}},
  \ and\ \bibinfo {author} {\bibfnamefont {C.}~\bibnamefont {Nayak}},\ }\href
  {\doibase 10.1038/npjqi.2015.1} {\bibfield  {journal} {\bibinfo  {journal}
  {Quantum Inf.}\ }\textbf {\bibinfo {volume} {1}},\ \bibinfo {pages} {15001}
  (\bibinfo {year} {2015})}\BibitemShut {NoStop}%
\bibitem [{\citenamefont {Fu}\ and\ \citenamefont {Kane}(2008)}]{Fu2008Mar}%
  \BibitemOpen
  \bibfield  {author} {\bibinfo {author} {\bibfnamefont {L.}~\bibnamefont
  {Fu}}\ and\ \bibinfo {author} {\bibfnamefont {C.~L.}\ \bibnamefont {Kane}},\
  }\href {\doibase 10.1103/PhysRevLett.100.096407} {\bibfield  {journal}
  {\bibinfo  {journal} {Phys. Rev. Lett.}\ }\textbf {\bibinfo {volume} {100}},\
  \bibinfo {pages} {096407} (\bibinfo {year} {2008})}\BibitemShut {NoStop}%
\bibitem [{\citenamefont {Lutchyn}\ \emph {et~al.}(2010)\citenamefont
  {Lutchyn}, \citenamefont {Sau},\ and\ \citenamefont
  {Das~Sarma}}]{Lutchyn2010Aug}%
  \BibitemOpen
  \bibfield  {author} {\bibinfo {author} {\bibfnamefont {R.~M.}\ \bibnamefont
  {Lutchyn}}, \bibinfo {author} {\bibfnamefont {J.~D.}\ \bibnamefont {Sau}}, \
  and\ \bibinfo {author} {\bibfnamefont {S.}~\bibnamefont {Das~Sarma}},\ }\href
  {\doibase 10.1103/PhysRevLett.105.077001} {\bibfield  {journal} {\bibinfo
  {journal} {Phys. Rev. Lett.}\ }\textbf {\bibinfo {volume} {105}},\ \bibinfo
  {pages} {077001} (\bibinfo {year} {2010})}\BibitemShut {NoStop}%
\bibitem [{\citenamefont {Oreg}\ \emph {et~al.}(2010)\citenamefont {Oreg},
  \citenamefont {Refael},\ and\ \citenamefont {von Oppen}}]{Oreg2010Oct}%
  \BibitemOpen
  \bibfield  {author} {\bibinfo {author} {\bibfnamefont {Y.}~\bibnamefont
  {Oreg}}, \bibinfo {author} {\bibfnamefont {G.}~\bibnamefont {Refael}}, \ and\
  \bibinfo {author} {\bibfnamefont {F.}~\bibnamefont {von Oppen}},\ }\href
  {\doibase 10.1103/PhysRevLett.105.177002} {\bibfield  {journal} {\bibinfo
  {journal} {Phys. Rev. Lett.}\ }\textbf {\bibinfo {volume} {105}},\ \bibinfo
  {pages} {177002} (\bibinfo {year} {2010})}\BibitemShut {NoStop}%
\bibitem [{\citenamefont {Cook}\ and\ \citenamefont
  {Franz}(2011)}]{Cook2011Nov}%
  \BibitemOpen
  \bibfield  {author} {\bibinfo {author} {\bibfnamefont {A.}~\bibnamefont
  {Cook}}\ and\ \bibinfo {author} {\bibfnamefont {M.}~\bibnamefont {Franz}},\
  }\href {\doibase 10.1103/PhysRevB.84.201105} {\bibfield  {journal} {\bibinfo
  {journal} {Phys. Rev. B}\ }\textbf {\bibinfo {volume} {84}},\ \bibinfo
  {pages} {201105(R)} (\bibinfo {year} {2011})}\BibitemShut {NoStop}%
\bibitem [{\citenamefont {Choy}\ \emph {et~al.}(2011)\citenamefont {Choy},
  \citenamefont {Edge}, \citenamefont {Akhmerov},\ and\ \citenamefont
  {Beenakker}}]{Choy2011Nov}%
  \BibitemOpen
  \bibfield  {author} {\bibinfo {author} {\bibfnamefont {T.-P.}\ \bibnamefont
  {Choy}}, \bibinfo {author} {\bibfnamefont {J.~M.}\ \bibnamefont {Edge}},
  \bibinfo {author} {\bibfnamefont {A.~R.}\ \bibnamefont {Akhmerov}}, \ and\
  \bibinfo {author} {\bibfnamefont {C.~W.~J.}\ \bibnamefont {Beenakker}},\
  }\href {\doibase 10.1103/PhysRevB.84.195442} {\bibfield  {journal} {\bibinfo
  {journal} {Phys. Rev. B}\ }\textbf {\bibinfo {volume} {84}},\ \bibinfo
  {pages} {195442} (\bibinfo {year} {2011})}\BibitemShut {NoStop}%
\bibitem [{\citenamefont {Vazifeh}\ and\ \citenamefont
  {Franz}(2013)}]{Vazifeh2013Nov}%
  \BibitemOpen
  \bibfield  {author} {\bibinfo {author} {\bibfnamefont {M.~M.}\ \bibnamefont
  {Vazifeh}}\ and\ \bibinfo {author} {\bibfnamefont {M.}~\bibnamefont
  {Franz}},\ }\href {\doibase 10.1103/PhysRevLett.111.206802} {\bibfield
  {journal} {\bibinfo  {journal} {Phys. Rev. Lett.}\ }\textbf {\bibinfo
  {volume} {111}},\ \bibinfo {pages} {206802} (\bibinfo {year}
  {2013})}\BibitemShut {NoStop}%
\bibitem [{\citenamefont {Klinovaja}\ \emph {et~al.}(2013)\citenamefont
  {Klinovaja}, \citenamefont {Stano}, \citenamefont {Yazdani},\ and\
  \citenamefont {Loss}}]{Klinovaja2013Nov}%
  \BibitemOpen
  \bibfield  {author} {\bibinfo {author} {\bibfnamefont {J.}~\bibnamefont
  {Klinovaja}}, \bibinfo {author} {\bibfnamefont {P.}~\bibnamefont {Stano}},
  \bibinfo {author} {\bibfnamefont {A.}~\bibnamefont {Yazdani}}, \ and\
  \bibinfo {author} {\bibfnamefont {D.}~\bibnamefont {Loss}},\ }\href {\doibase
  10.1103/PhysRevLett.111.186805} {\bibfield  {journal} {\bibinfo  {journal}
  {Phys. Rev. Lett.}\ }\textbf {\bibinfo {volume} {111}},\ \bibinfo {pages}
  {186805} (\bibinfo {year} {2013})}\BibitemShut {NoStop}%
\bibitem [{\citenamefont {Beenakker}(2013)}]{Beenakker2013Mar}%
  \BibitemOpen
  \bibfield  {author} {\bibinfo {author} {\bibfnamefont {C.~W.~J.}\
  \bibnamefont {Beenakker}},\ }\href {\doibase
  10.1146/annurev-conmatphys-030212-184337} {\bibfield  {journal} {\bibinfo
  {journal} {Annu. Rev. Condens. Matter Phys.}\ }\textbf {\bibinfo {volume}
  {4}},\ \bibinfo {pages} {113} (\bibinfo {year} {2013})}\BibitemShut {NoStop}%
\bibitem [{\citenamefont {Nadj-Perge}\ \emph {et~al.}(2014)\citenamefont
  {Nadj-Perge}, \citenamefont {Drozdov}, \citenamefont {Li}, \citenamefont
  {Chen}, \citenamefont {Jeon}, \citenamefont {Seo}, \citenamefont {MacDonald},
  \citenamefont {Bernevig},\ and\ \citenamefont {Yazdani}}]{Nadj-Perge2014Oct}%
  \BibitemOpen
  \bibfield  {author} {\bibinfo {author} {\bibfnamefont {S.}~\bibnamefont
  {Nadj-Perge}}, \bibinfo {author} {\bibfnamefont {I.~K.}\ \bibnamefont
  {Drozdov}}, \bibinfo {author} {\bibfnamefont {J.}~\bibnamefont {Li}},
  \bibinfo {author} {\bibfnamefont {H.}~\bibnamefont {Chen}}, \bibinfo {author}
  {\bibfnamefont {S.}~\bibnamefont {Jeon}}, \bibinfo {author} {\bibfnamefont
  {J.}~\bibnamefont {Seo}}, \bibinfo {author} {\bibfnamefont {A.~H.}\
  \bibnamefont {MacDonald}}, \bibinfo {author} {\bibfnamefont {B.~A.}\
  \bibnamefont {Bernevig}}, \ and\ \bibinfo {author} {\bibfnamefont
  {A.}~\bibnamefont {Yazdani}},\ }\href {\doibase 10.1126/science.1259327}
  {\bibfield  {journal} {\bibinfo  {journal} {Science}\ }\textbf {\bibinfo
  {volume} {346}},\ \bibinfo {pages} {602} (\bibinfo {year}
  {2014})}\BibitemShut {NoStop}%
\bibitem [{\citenamefont {Elliott}\ and\ \citenamefont
  {Franz}(2015)}]{Elliott2015Feb}%
  \BibitemOpen
  \bibfield  {author} {\bibinfo {author} {\bibfnamefont {S.~R.}\ \bibnamefont
  {Elliott}}\ and\ \bibinfo {author} {\bibfnamefont {M.}~\bibnamefont
  {Franz}},\ }\href {\doibase 10.1103/RevModPhys.87.137} {\bibfield  {journal}
  {\bibinfo  {journal} {Rev. Mod. Phys.}\ }\textbf {\bibinfo {volume} {87}},\
  \bibinfo {pages} {137} (\bibinfo {year} {2015})}\BibitemShut {NoStop}%
\bibitem [{\citenamefont {Lutchyn}\ \emph {et~al.}(2018)\citenamefont
  {Lutchyn}, \citenamefont {Bakkers}, \citenamefont {Kouwenhoven},
  \citenamefont {Krogstrup}, \citenamefont {Marcus},\ and\ \citenamefont
  {Oreg}}]{Lutchyn2018May}%
  \BibitemOpen
  \bibfield  {author} {\bibinfo {author} {\bibfnamefont {R.~M.}\ \bibnamefont
  {Lutchyn}}, \bibinfo {author} {\bibfnamefont {E.~P. A.~M.}\ \bibnamefont
  {Bakkers}}, \bibinfo {author} {\bibfnamefont {L.~P.}\ \bibnamefont
  {Kouwenhoven}}, \bibinfo {author} {\bibfnamefont {P.}~\bibnamefont
  {Krogstrup}}, \bibinfo {author} {\bibfnamefont {C.~M.}\ \bibnamefont
  {Marcus}}, \ and\ \bibinfo {author} {\bibfnamefont {Y.}~\bibnamefont
  {Oreg}},\ }\href {\doibase 10.1038/s41578-018-0003-1} {\bibfield  {journal}
  {\bibinfo  {journal} {Nat. Rev. Mater.}\ }\textbf {\bibinfo {volume} {3}},\
  \bibinfo {pages} {52} (\bibinfo {year} {2018})}\BibitemShut {NoStop}%
\bibitem [{\citenamefont {Fornieri}\ \emph {et~al.}(2019)\citenamefont
  {Fornieri}, \citenamefont {Whiticar}, \citenamefont {Setiawan}, \citenamefont
  {Portol{\ifmmode\acute{e}\else\'{e}\fi}s}, \citenamefont {Drachmann},
  \citenamefont {Keselman}, \citenamefont {Gronin}, \citenamefont {Thomas},
  \citenamefont {Wang}, \citenamefont {Kallaher}, \citenamefont {Gardner},
  \citenamefont {Berg}, \citenamefont {Manfra}, \citenamefont {Stern},
  \citenamefont {Marcus},\ and\ \citenamefont {Nichele}}]{Fornieri2019Apr}%
  \BibitemOpen
  \bibfield  {author} {\bibinfo {author} {\bibfnamefont {A.}~\bibnamefont
  {Fornieri}}, \bibinfo {author} {\bibfnamefont {A.~M.}\ \bibnamefont
  {Whiticar}}, \bibinfo {author} {\bibfnamefont {F.}~\bibnamefont {Setiawan}},
  \bibinfo {author} {\bibfnamefont {E.}~\bibnamefont
  {Portol{\ifmmode\acute{e}\else\'{e}\fi}s}}, \bibinfo {author} {\bibfnamefont
  {A.~C.~C.}\ \bibnamefont {Drachmann}}, \bibinfo {author} {\bibfnamefont
  {A.}~\bibnamefont {Keselman}}, \bibinfo {author} {\bibfnamefont
  {S.}~\bibnamefont {Gronin}}, \bibinfo {author} {\bibfnamefont
  {C.}~\bibnamefont {Thomas}}, \bibinfo {author} {\bibfnamefont
  {T.}~\bibnamefont {Wang}}, \bibinfo {author} {\bibfnamefont {R.}~\bibnamefont
  {Kallaher}}, \bibinfo {author} {\bibfnamefont {G.~C.}\ \bibnamefont
  {Gardner}}, \bibinfo {author} {\bibfnamefont {E.}~\bibnamefont {Berg}},
  \bibinfo {author} {\bibfnamefont {M.~J.}\ \bibnamefont {Manfra}}, \bibinfo
  {author} {\bibfnamefont {A.}~\bibnamefont {Stern}}, \bibinfo {author}
  {\bibfnamefont {C.~M.}\ \bibnamefont {Marcus}}, \ and\ \bibinfo {author}
  {\bibfnamefont {F.}~\bibnamefont {Nichele}},\ }\href {\doibase
  10.1038/s41586-019-1068-8} {\bibfield  {journal} {\bibinfo  {journal}
  {Nature}\ }\textbf {\bibinfo {volume} {569}},\ \bibinfo {pages} {89}
  (\bibinfo {year} {2019})}\BibitemShut {NoStop}%
\bibitem [{\citenamefont {Kitaev}(2001)}]{Kitaev2001}%
  \BibitemOpen
  \bibfield  {author} {\bibinfo {author} {\bibfnamefont {A.~Y.}\ \bibnamefont
  {Kitaev}},\ }\href {\doibase 10.1070/1063-7869/44/10s/s29} {\bibfield
  {journal} {\bibinfo  {journal} {Phys. Usp.}\ }\textbf {\bibinfo {volume}
  {44}},\ \bibinfo {pages} {131} (\bibinfo {year} {2001})}\BibitemShut
  {NoStop}%
\bibitem [{\citenamefont {Mourik}\ \emph {et~al.}(2012)\citenamefont {Mourik},
  \citenamefont {Zuo}, \citenamefont {Frolov}, \citenamefont {Plissard},
  \citenamefont {Bakkers},\ and\ \citenamefont {Kouwenhoven}}]{Mourik2012May}%
  \BibitemOpen
  \bibfield  {author} {\bibinfo {author} {\bibfnamefont {V.}~\bibnamefont
  {Mourik}}, \bibinfo {author} {\bibfnamefont {K.}~\bibnamefont {Zuo}},
  \bibinfo {author} {\bibfnamefont {S.~M.}\ \bibnamefont {Frolov}}, \bibinfo
  {author} {\bibfnamefont {S.~R.}\ \bibnamefont {Plissard}}, \bibinfo {author}
  {\bibfnamefont {E.~P. A.~M.}\ \bibnamefont {Bakkers}}, \ and\ \bibinfo
  {author} {\bibfnamefont {L.~P.}\ \bibnamefont {Kouwenhoven}},\ }\href
  {\doibase 10.1126/science.1222360} {\bibfield  {journal} {\bibinfo  {journal}
  {Science}\ }\textbf {\bibinfo {volume} {336}},\ \bibinfo {pages} {1003}
  (\bibinfo {year} {2012})}\BibitemShut {NoStop}%
\bibitem [{\citenamefont {Das}\ \emph {et~al.}(2012)\citenamefont {Das},
  \citenamefont {Ronen}, \citenamefont {Most}, \citenamefont {Oreg},
  \citenamefont {Heiblum},\ and\ \citenamefont {Shtrikman}}]{Das2012Nov}%
  \BibitemOpen
  \bibfield  {author} {\bibinfo {author} {\bibfnamefont {A.}~\bibnamefont
  {Das}}, \bibinfo {author} {\bibfnamefont {Y.}~\bibnamefont {Ronen}}, \bibinfo
  {author} {\bibfnamefont {Y.}~\bibnamefont {Most}}, \bibinfo {author}
  {\bibfnamefont {Y.}~\bibnamefont {Oreg}}, \bibinfo {author} {\bibfnamefont
  {M.}~\bibnamefont {Heiblum}}, \ and\ \bibinfo {author} {\bibfnamefont
  {H.}~\bibnamefont {Shtrikman}},\ }\href {\doibase 10.1038/nphys2479}
  {\bibfield  {journal} {\bibinfo  {journal} {Nat. Phys.}\ }\textbf {\bibinfo
  {volume} {8}},\ \bibinfo {pages} {887} (\bibinfo {year} {2012})}\BibitemShut
  {NoStop}%
\bibitem [{\citenamefont {Rokhinson}\ \emph {et~al.}(2012)\citenamefont
  {Rokhinson}, \citenamefont {Liu},\ and\ \citenamefont
  {Furdyna}}]{Rokhinson2012Sep}%
  \BibitemOpen
  \bibfield  {author} {\bibinfo {author} {\bibfnamefont {L.~P.}\ \bibnamefont
  {Rokhinson}}, \bibinfo {author} {\bibfnamefont {X.}~\bibnamefont {Liu}}, \
  and\ \bibinfo {author} {\bibfnamefont {J.~K.}\ \bibnamefont {Furdyna}},\
  }\href {\doibase 10.1038/nphys2429} {\bibfield  {journal} {\bibinfo
  {journal} {Nat. Phys.}\ }\textbf {\bibinfo {volume} {8}},\ \bibinfo {pages}
  {795} (\bibinfo {year} {2012})}\BibitemShut {NoStop}%
\bibitem [{\citenamefont {Wiedenmann}\ \emph {et~al.}(2016)\citenamefont
  {Wiedenmann}, \citenamefont {Bocquillon}, \citenamefont {Deacon},
  \citenamefont {Hartinger}, \citenamefont {Herrmann}, \citenamefont
  {Klapwijk}, \citenamefont {Maier}, \citenamefont {Ames}, \citenamefont
  {Br{\ifmmode\ddot{u}\else\"{u}\fi}ne}, \citenamefont {Gould}, \citenamefont
  {Oiwa}, \citenamefont {Ishibashi}, \citenamefont {Tarucha}, \citenamefont
  {Buhmann},\ and\ \citenamefont {Molenkamp}}]{Wiedenmann2016Jan}%
  \BibitemOpen
  \bibfield  {author} {\bibinfo {author} {\bibfnamefont {J.}~\bibnamefont
  {Wiedenmann}}, \bibinfo {author} {\bibfnamefont {E.}~\bibnamefont
  {Bocquillon}}, \bibinfo {author} {\bibfnamefont {R.~S.}\ \bibnamefont
  {Deacon}}, \bibinfo {author} {\bibfnamefont {S.}~\bibnamefont {Hartinger}},
  \bibinfo {author} {\bibfnamefont {O.}~\bibnamefont {Herrmann}}, \bibinfo
  {author} {\bibfnamefont {T.~M.}\ \bibnamefont {Klapwijk}}, \bibinfo {author}
  {\bibfnamefont {L.}~\bibnamefont {Maier}}, \bibinfo {author} {\bibfnamefont
  {C.}~\bibnamefont {Ames}}, \bibinfo {author} {\bibfnamefont {C.}~\bibnamefont
  {Br{\ifmmode\ddot{u}\else\"{u}\fi}ne}}, \bibinfo {author} {\bibfnamefont
  {C.}~\bibnamefont {Gould}}, \bibinfo {author} {\bibfnamefont
  {A.}~\bibnamefont {Oiwa}}, \bibinfo {author} {\bibfnamefont {K.}~\bibnamefont
  {Ishibashi}}, \bibinfo {author} {\bibfnamefont {S.}~\bibnamefont {Tarucha}},
  \bibinfo {author} {\bibfnamefont {H.}~\bibnamefont {Buhmann}}, \ and\
  \bibinfo {author} {\bibfnamefont {L.~W.}\ \bibnamefont {Molenkamp}},\ }\href
  {\doibase 10.1038/ncomms10303} {\bibfield  {journal} {\bibinfo  {journal}
  {Nat. Commun.}\ }\textbf {\bibinfo {volume} {7}},\ \bibinfo {pages} {10303}
  (\bibinfo {year} {2016})}\BibitemShut {NoStop}%
\bibitem [{\citenamefont {Deng}\ \emph {et~al.}(2016)\citenamefont {Deng},
  \citenamefont {Vaitiek{\ifmmode\dot{e}\else\.{e}\fi}nas}, \citenamefont
  {Hansen}, \citenamefont {Danon}, \citenamefont {Leijnse}, \citenamefont
  {Flensberg}, \citenamefont {Nyg{\aa}rd}, \citenamefont {Krogstrup},\ and\
  \citenamefont {Marcus}}]{Deng2016Dec}%
  \BibitemOpen
  \bibfield  {author} {\bibinfo {author} {\bibfnamefont {M.~T.}\ \bibnamefont
  {Deng}}, \bibinfo {author} {\bibfnamefont {S.}~\bibnamefont
  {Vaitiek{\ifmmode\dot{e}\else\.{e}\fi}nas}}, \bibinfo {author} {\bibfnamefont
  {E.~B.}\ \bibnamefont {Hansen}}, \bibinfo {author} {\bibfnamefont
  {J.}~\bibnamefont {Danon}}, \bibinfo {author} {\bibfnamefont
  {M.}~\bibnamefont {Leijnse}}, \bibinfo {author} {\bibfnamefont
  {K.}~\bibnamefont {Flensberg}}, \bibinfo {author} {\bibfnamefont
  {J.}~\bibnamefont {Nyg{\aa}rd}}, \bibinfo {author} {\bibfnamefont
  {P.}~\bibnamefont {Krogstrup}}, \ and\ \bibinfo {author} {\bibfnamefont
  {C.~M.}\ \bibnamefont {Marcus}},\ }\href {\doibase 10.1126/science.aaf3961}
  {\bibfield  {journal} {\bibinfo  {journal} {Science}\ }\textbf {\bibinfo
  {volume} {354}},\ \bibinfo {pages} {1557} (\bibinfo {year}
  {2016})}\BibitemShut {NoStop}%
\bibitem [{\citenamefont {Aguado}(2017)}]{Aguado2017Oct}%
  \BibitemOpen
  \bibfield  {author} {\bibinfo {author} {\bibfnamefont {R.}~\bibnamefont
  {Aguado}},\ }\href {\doibase 10.1393/ncr/i2017-10141-9} {\bibfield  {journal}
  {\bibinfo  {journal} {Riv. Nuovo Cimento}\ }\textbf {\bibinfo {volume}
  {11}},\ \bibinfo {pages} {523} (\bibinfo {year} {2017})}\BibitemShut
  {NoStop}%
\bibitem [{\citenamefont {Zhang}\ \emph {et~al.}(2018)\citenamefont {Zhang},
  \citenamefont {Liu}, \citenamefont {Gazibegovic}, \citenamefont {Xu},
  \citenamefont {Logan}, \citenamefont {Wang}, \citenamefont {van Loo},
  \citenamefont {Bommer}, \citenamefont {de~Moor}, \citenamefont {Car},
  \citenamefont {Op~het Veld}, \citenamefont {van Veldhoven}, \citenamefont
  {Koelling}, \citenamefont {Verheijen}, \citenamefont {Pendharkar},
  \citenamefont {Pennachio}, \citenamefont {Shojaei}, \citenamefont {Lee},
  \citenamefont {Palmstr{\o}m}, \citenamefont {Bakkers}, \citenamefont
  {Sarma},\ and\ \citenamefont {Kouwenhoven}}]{Zhang2018Mar}%
  \BibitemOpen
  \bibfield  {author} {\bibinfo {author} {\bibfnamefont {H.}~\bibnamefont
  {Zhang}}, \bibinfo {author} {\bibfnamefont {C.-X.}\ \bibnamefont {Liu}},
  \bibinfo {author} {\bibfnamefont {S.}~\bibnamefont {Gazibegovic}}, \bibinfo
  {author} {\bibfnamefont {D.}~\bibnamefont {Xu}}, \bibinfo {author}
  {\bibfnamefont {J.~A.}\ \bibnamefont {Logan}}, \bibinfo {author}
  {\bibfnamefont {G.}~\bibnamefont {Wang}}, \bibinfo {author} {\bibfnamefont
  {N.}~\bibnamefont {van Loo}}, \bibinfo {author} {\bibfnamefont {J.~D.~S.}\
  \bibnamefont {Bommer}}, \bibinfo {author} {\bibfnamefont {M.~W.~A.}\
  \bibnamefont {de~Moor}}, \bibinfo {author} {\bibfnamefont {D.}~\bibnamefont
  {Car}}, \bibinfo {author} {\bibfnamefont {R.~L.~M.}\ \bibnamefont {Op~het
  Veld}}, \bibinfo {author} {\bibfnamefont {P.~J.}\ \bibnamefont {van
  Veldhoven}}, \bibinfo {author} {\bibfnamefont {S.}~\bibnamefont {Koelling}},
  \bibinfo {author} {\bibfnamefont {M.~A.}\ \bibnamefont {Verheijen}}, \bibinfo
  {author} {\bibfnamefont {M.}~\bibnamefont {Pendharkar}}, \bibinfo {author}
  {\bibfnamefont {D.~J.}\ \bibnamefont {Pennachio}}, \bibinfo {author}
  {\bibfnamefont {B.}~\bibnamefont {Shojaei}}, \bibinfo {author} {\bibfnamefont
  {J.~S.}\ \bibnamefont {Lee}}, \bibinfo {author} {\bibfnamefont {C.~J.}\
  \bibnamefont {Palmstr{\o}m}}, \bibinfo {author} {\bibfnamefont {E.~P. A.~M.}\
  \bibnamefont {Bakkers}}, \bibinfo {author} {\bibfnamefont {S.~D.}\
  \bibnamefont {Sarma}}, \ and\ \bibinfo {author} {\bibfnamefont {L.~P.}\
  \bibnamefont {Kouwenhoven}},\ }\href {\doibase 10.1038/nature26142}
  {\bibfield  {journal} {\bibinfo  {journal} {Nature}\ }\textbf {\bibinfo
  {volume} {556}},\ \bibinfo {pages} {74} (\bibinfo {year} {2018})}\BibitemShut
  {NoStop}%
\bibitem [{\citenamefont {Akhmerov}\ \emph {et~al.}(2011)\citenamefont
  {Akhmerov}, \citenamefont {Dahlhaus}, \citenamefont {Hassler}, \citenamefont
  {Wimmer},\ and\ \citenamefont {Beenakker}}]{Akhmerov2011Jan}%
  \BibitemOpen
  \bibfield  {author} {\bibinfo {author} {\bibfnamefont {A.~R.}\ \bibnamefont
  {Akhmerov}}, \bibinfo {author} {\bibfnamefont {J.~P.}\ \bibnamefont
  {Dahlhaus}}, \bibinfo {author} {\bibfnamefont {F.}~\bibnamefont {Hassler}},
  \bibinfo {author} {\bibfnamefont {M.}~\bibnamefont {Wimmer}}, \ and\ \bibinfo
  {author} {\bibfnamefont {C.~W.~J.}\ \bibnamefont {Beenakker}},\ }\href
  {\doibase 10.1103/PhysRevLett.106.057001} {\bibfield  {journal} {\bibinfo
  {journal} {Phys. Rev. Lett.}\ }\textbf {\bibinfo {volume} {106}},\ \bibinfo
  {pages} {057001} (\bibinfo {year} {2011})}\BibitemShut {NoStop}%
\bibitem [{\citenamefont {Lee}\ \emph {et~al.}(2013{\natexlab{a}})\citenamefont
  {Lee}, \citenamefont {Jiang}, \citenamefont {Houzet}, \citenamefont {Aguado},
  \citenamefont {Lieber},\ and\ \citenamefont {De~Franceschi}}]{Lee2013Dec}%
  \BibitemOpen
  \bibfield  {author} {\bibinfo {author} {\bibfnamefont {E.~J.~H.}\
  \bibnamefont {Lee}}, \bibinfo {author} {\bibfnamefont {X.}~\bibnamefont
  {Jiang}}, \bibinfo {author} {\bibfnamefont {M.}~\bibnamefont {Houzet}},
  \bibinfo {author} {\bibfnamefont {R.}~\bibnamefont {Aguado}}, \bibinfo
  {author} {\bibfnamefont {C.~M.}\ \bibnamefont {Lieber}}, \ and\ \bibinfo
  {author} {\bibfnamefont {S.}~\bibnamefont {De~Franceschi}},\ }\href {\doibase
  10.1038/nnano.2013.267} {\bibfield  {journal} {\bibinfo  {journal} {Nat.
  Nanotechnol.}\ }\textbf {\bibinfo {volume} {9}},\ \bibinfo {pages} {79}
  (\bibinfo {year} {2013}{\natexlab{a}})}\BibitemShut {NoStop}%
\bibitem [{\citenamefont {Cayao}\ \emph {et~al.}(2015)\citenamefont {Cayao},
  \citenamefont {Prada}, \citenamefont {San-Jose},\ and\ \citenamefont
  {Aguado}}]{Cayao2015Jan}%
  \BibitemOpen
  \bibfield  {author} {\bibinfo {author} {\bibfnamefont {J.}~\bibnamefont
  {Cayao}}, \bibinfo {author} {\bibfnamefont {E.}~\bibnamefont {Prada}},
  \bibinfo {author} {\bibfnamefont {P.}~\bibnamefont {San-Jose}}, \ and\
  \bibinfo {author} {\bibfnamefont {R.}~\bibnamefont {Aguado}},\ }\href
  {\doibase 10.1103/PhysRevB.91.024514} {\bibfield  {journal} {\bibinfo
  {journal} {Phys. Rev. B}\ }\textbf {\bibinfo {volume} {91}},\ \bibinfo
  {pages} {024514} (\bibinfo {year} {2015})}\BibitemShut {NoStop}%
\bibitem [{\citenamefont {San-Jose}\ \emph {et~al.}(2016)\citenamefont
  {San-Jose}, \citenamefont {Cayao}, \citenamefont {Prada},\ and\ \citenamefont
  {Aguado}}]{San-Jose2016Feb}%
  \BibitemOpen
  \bibfield  {author} {\bibinfo {author} {\bibfnamefont {P.}~\bibnamefont
  {San-Jose}}, \bibinfo {author} {\bibfnamefont {J.}~\bibnamefont {Cayao}},
  \bibinfo {author} {\bibfnamefont {E.}~\bibnamefont {Prada}}, \ and\ \bibinfo
  {author} {\bibfnamefont {R.}~\bibnamefont {Aguado}},\ }\href {\doibase
  10.1038/srep21427} {\bibfield  {journal} {\bibinfo  {journal} {Sci. Rep.}\
  }\textbf {\bibinfo {volume} {6}},\ \bibinfo {pages} {21427} (\bibinfo {year}
  {2016})}\BibitemShut {NoStop}%
\bibitem [{\citenamefont {Liu}\ \emph {et~al.}(2017)\citenamefont {Liu},
  \citenamefont {Sau}, \citenamefont {Stanescu},\ and\ \citenamefont
  {Das~Sarma}}]{Liu2017Aug}%
  \BibitemOpen
  \bibfield  {author} {\bibinfo {author} {\bibfnamefont {C.-X.}\ \bibnamefont
  {Liu}}, \bibinfo {author} {\bibfnamefont {J.~D.}\ \bibnamefont {Sau}},
  \bibinfo {author} {\bibfnamefont {T.~D.}\ \bibnamefont {Stanescu}}, \ and\
  \bibinfo {author} {\bibfnamefont {S.}~\bibnamefont {Das~Sarma}},\ }\href
  {\doibase 10.1103/PhysRevB.96.075161} {\bibfield  {journal} {\bibinfo
  {journal} {Phys. Rev. B}\ }\textbf {\bibinfo {volume} {96}},\ \bibinfo
  {pages} {075161} (\bibinfo {year} {2017})}\BibitemShut {NoStop}%
\bibitem [{\citenamefont {Liu}\ \emph {et~al.}(2018)\citenamefont {Liu},
  \citenamefont {Sau},\ and\ \citenamefont {Das~Sarma}}]{Liu2018Jun}%
  \BibitemOpen
  \bibfield  {author} {\bibinfo {author} {\bibfnamefont {C.-X.}\ \bibnamefont
  {Liu}}, \bibinfo {author} {\bibfnamefont {J.~D.}\ \bibnamefont {Sau}}, \ and\
  \bibinfo {author} {\bibfnamefont {S.}~\bibnamefont {Das~Sarma}},\ }\href
  {\doibase 10.1103/PhysRevB.97.214502} {\bibfield  {journal} {\bibinfo
  {journal} {Phys. Rev. B}\ }\textbf {\bibinfo {volume} {97}},\ \bibinfo
  {pages} {214502} (\bibinfo {year} {2018})}\BibitemShut {NoStop}%
\bibitem [{\citenamefont {Hell}\ \emph {et~al.}(2018)\citenamefont {Hell},
  \citenamefont {Flensberg},\ and\ \citenamefont {Leijnse}}]{Hell2018Apr}%
  \BibitemOpen
  \bibfield  {author} {\bibinfo {author} {\bibfnamefont {M.}~\bibnamefont
  {Hell}}, \bibinfo {author} {\bibfnamefont {K.}~\bibnamefont {Flensberg}}, \
  and\ \bibinfo {author} {\bibfnamefont {M.}~\bibnamefont {Leijnse}},\ }\href
  {\doibase 10.1103/PhysRevB.97.161401} {\bibfield  {journal} {\bibinfo
  {journal} {Phys. Rev. B}\ }\textbf {\bibinfo {volume} {97}},\ \bibinfo
  {pages} {161401(R)} (\bibinfo {year} {2018})}\BibitemShut {NoStop}%
\bibitem [{\citenamefont {Vuik}\ \emph {et~al.}(2019)\citenamefont {Vuik},
  \citenamefont {Nijholt}, \citenamefont {Akhmerov},\ and\ \citenamefont
  {Wimmer}}]{Vuik2019Nov}%
  \BibitemOpen
  \bibfield  {author} {\bibinfo {author} {\bibfnamefont {A.}~\bibnamefont
  {Vuik}}, \bibinfo {author} {\bibfnamefont {B.}~\bibnamefont {Nijholt}},
  \bibinfo {author} {\bibfnamefont {A.}~\bibnamefont {Akhmerov}}, \ and\
  \bibinfo {author} {\bibfnamefont {M.}~\bibnamefont {Wimmer}},\ }\href
  {\doibase 10.21468/SciPostPhys.7.5.061} {\bibfield  {journal} {\bibinfo
  {journal} {SciPost Physics}\ }\textbf {\bibinfo {volume} {7}},\ \bibinfo
  {pages} {061} (\bibinfo {year} {2019})}\BibitemShut {NoStop}%
\bibitem [{\citenamefont {Chiu}\ and\ \citenamefont
  {Das~Sarma}(2019)}]{Chiu2019Jan}%
  \BibitemOpen
  \bibfield  {author} {\bibinfo {author} {\bibfnamefont {C.-K.}\ \bibnamefont
  {Chiu}}\ and\ \bibinfo {author} {\bibfnamefont {S.}~\bibnamefont
  {Das~Sarma}},\ }\href {\doibase 10.1103/PhysRevB.99.035312} {\bibfield
  {journal} {\bibinfo  {journal} {Phys. Rev. B}\ }\textbf {\bibinfo {volume}
  {99}},\ \bibinfo {pages} {035312} (\bibinfo {year} {2019})}\BibitemShut
  {NoStop}%
\bibitem [{\citenamefont {Read}\ and\ \citenamefont
  {Green}(2000)}]{Read2000Apr}%
  \BibitemOpen
  \bibfield  {author} {\bibinfo {author} {\bibfnamefont {N.}~\bibnamefont
  {Read}}\ and\ \bibinfo {author} {\bibfnamefont {D.}~\bibnamefont {Green}},\
  }\href {\doibase 10.1103/PhysRevB.61.10267} {\bibfield  {journal} {\bibinfo
  {journal} {Phys. Rev. B}\ }\textbf {\bibinfo {volume} {61}},\ \bibinfo
  {pages} {10267} (\bibinfo {year} {2000})}\BibitemShut {NoStop}%
\bibitem [{\citenamefont {Ivanov}(2001)}]{Ivanov2001Jan}%
  \BibitemOpen
  \bibfield  {author} {\bibinfo {author} {\bibfnamefont {D.~A.}\ \bibnamefont
  {Ivanov}},\ }\href {\doibase 10.1103/PhysRevLett.86.268} {\bibfield
  {journal} {\bibinfo  {journal} {Phys. Rev. Lett.}\ }\textbf {\bibinfo
  {volume} {86}},\ \bibinfo {pages} {268} (\bibinfo {year} {2001})}\BibitemShut
  {NoStop}%
\bibitem [{\citenamefont {Alicea}\ \emph {et~al.}(2011)\citenamefont {Alicea},
  \citenamefont {Oreg}, \citenamefont {Refael}, \citenamefont {von Oppen},\
  and\ \citenamefont {Fisher}}]{Alicea2011Feb}%
  \BibitemOpen
  \bibfield  {author} {\bibinfo {author} {\bibfnamefont {J.}~\bibnamefont
  {Alicea}}, \bibinfo {author} {\bibfnamefont {Y.}~\bibnamefont {Oreg}},
  \bibinfo {author} {\bibfnamefont {G.}~\bibnamefont {Refael}}, \bibinfo
  {author} {\bibfnamefont {F.}~\bibnamefont {von Oppen}}, \ and\ \bibinfo
  {author} {\bibfnamefont {M.~P.~A.}\ \bibnamefont {Fisher}},\ }\href {\doibase
  10.1038/nphys1915} {\bibfield  {journal} {\bibinfo  {journal} {Nat. Phys.}\
  }\textbf {\bibinfo {volume} {7}},\ \bibinfo {pages} {412} (\bibinfo {year}
  {2011})}\BibitemShut {NoStop}%
\bibitem [{\citenamefont {Flensberg}(2011)}]{Flensberg2011Mar}%
  \BibitemOpen
  \bibfield  {author} {\bibinfo {author} {\bibfnamefont {K.}~\bibnamefont
  {Flensberg}},\ }\href {\doibase 10.1103/PhysRevLett.106.090503} {\bibfield
  {journal} {\bibinfo  {journal} {Phys. Rev. Lett.}\ }\textbf {\bibinfo
  {volume} {106}},\ \bibinfo {pages} {090503} (\bibinfo {year}
  {2011})}\BibitemShut {NoStop}%
\bibitem [{\citenamefont {Sau}\ \emph {et~al.}(2011)\citenamefont {Sau},
  \citenamefont {Clarke},\ and\ \citenamefont {Tewari}}]{Sau2011Sep}%
  \BibitemOpen
  \bibfield  {author} {\bibinfo {author} {\bibfnamefont {J.~D.}\ \bibnamefont
  {Sau}}, \bibinfo {author} {\bibfnamefont {D.~J.}\ \bibnamefont {Clarke}}, \
  and\ \bibinfo {author} {\bibfnamefont {S.}~\bibnamefont {Tewari}},\ }\href
  {\doibase 10.1103/PhysRevB.84.094505} {\bibfield  {journal} {\bibinfo
  {journal} {Phys. Rev. B}\ }\textbf {\bibinfo {volume} {84}},\ \bibinfo
  {pages} {094505} (\bibinfo {year} {2011})}\BibitemShut {NoStop}%
\bibitem [{\citenamefont {van Heck}\ \emph {et~al.}(2012)\citenamefont {van
  Heck}, \citenamefont {Akhmerov}, \citenamefont {Hassler}, \citenamefont
  {Burrello},\ and\ \citenamefont {Beenakker}}]{vanHeck2012Mar}%
  \BibitemOpen
  \bibfield  {author} {\bibinfo {author} {\bibfnamefont {B.}~\bibnamefont {van
  Heck}}, \bibinfo {author} {\bibfnamefont {A.~R.}\ \bibnamefont {Akhmerov}},
  \bibinfo {author} {\bibfnamefont {F.}~\bibnamefont {Hassler}}, \bibinfo
  {author} {\bibfnamefont {M.}~\bibnamefont {Burrello}}, \ and\ \bibinfo
  {author} {\bibfnamefont {C.~W.~J.}\ \bibnamefont {Beenakker}},\ }\href
  {\doibase 10.1088/1367-2630/14/3/035019} {\bibfield  {journal} {\bibinfo
  {journal} {New J. Phys.}\ }\textbf {\bibinfo {volume} {14}},\ \bibinfo
  {pages} {035019} (\bibinfo {year} {2012})}\BibitemShut {NoStop}%
\bibitem [{\citenamefont {Aasen}\ \emph {et~al.}(2016)\citenamefont {Aasen},
  \citenamefont {Hell}, \citenamefont {Mishmash}, \citenamefont {Higginbotham},
  \citenamefont {Danon}, \citenamefont {Leijnse}, \citenamefont {Jespersen},
  \citenamefont {Folk}, \citenamefont {Marcus}, \citenamefont {Flensberg},\
  and\ \citenamefont {Alicea}}]{Aasen2016Aug}%
  \BibitemOpen
  \bibfield  {author} {\bibinfo {author} {\bibfnamefont {D.}~\bibnamefont
  {Aasen}}, \bibinfo {author} {\bibfnamefont {M.}~\bibnamefont {Hell}},
  \bibinfo {author} {\bibfnamefont {R.~V.}\ \bibnamefont {Mishmash}}, \bibinfo
  {author} {\bibfnamefont {A.}~\bibnamefont {Higginbotham}}, \bibinfo {author}
  {\bibfnamefont {J.}~\bibnamefont {Danon}}, \bibinfo {author} {\bibfnamefont
  {M.}~\bibnamefont {Leijnse}}, \bibinfo {author} {\bibfnamefont {T.~S.}\
  \bibnamefont {Jespersen}}, \bibinfo {author} {\bibfnamefont {J.~A.}\
  \bibnamefont {Folk}}, \bibinfo {author} {\bibfnamefont {C.~M.}\ \bibnamefont
  {Marcus}}, \bibinfo {author} {\bibfnamefont {K.}~\bibnamefont {Flensberg}}, \
  and\ \bibinfo {author} {\bibfnamefont {J.}~\bibnamefont {Alicea}},\ }\href
  {\doibase 10.1103/PhysRevX.6.031016} {\bibfield  {journal} {\bibinfo
  {journal} {Phys. Rev. X}\ }\textbf {\bibinfo {volume} {6}},\ \bibinfo {pages}
  {031016} (\bibinfo {year} {2016})}\BibitemShut {NoStop}%
\bibitem [{\citenamefont {Karzig}\ \emph {et~al.}(2017)\citenamefont {Karzig},
  \citenamefont {Knapp}, \citenamefont {Lutchyn}, \citenamefont {Bonderson},
  \citenamefont {Hastings}, \citenamefont {Nayak}, \citenamefont {Alicea},
  \citenamefont {Flensberg}, \citenamefont {Plugge}, \citenamefont {Oreg},
  \citenamefont {Marcus},\ and\ \citenamefont {Freedman}}]{Karzig2017Jun}%
  \BibitemOpen
  \bibfield  {author} {\bibinfo {author} {\bibfnamefont {T.}~\bibnamefont
  {Karzig}}, \bibinfo {author} {\bibfnamefont {C.}~\bibnamefont {Knapp}},
  \bibinfo {author} {\bibfnamefont {R.~M.}\ \bibnamefont {Lutchyn}}, \bibinfo
  {author} {\bibfnamefont {P.}~\bibnamefont {Bonderson}}, \bibinfo {author}
  {\bibfnamefont {M.~B.}\ \bibnamefont {Hastings}}, \bibinfo {author}
  {\bibfnamefont {C.}~\bibnamefont {Nayak}}, \bibinfo {author} {\bibfnamefont
  {J.}~\bibnamefont {Alicea}}, \bibinfo {author} {\bibfnamefont
  {K.}~\bibnamefont {Flensberg}}, \bibinfo {author} {\bibfnamefont
  {S.}~\bibnamefont {Plugge}}, \bibinfo {author} {\bibfnamefont
  {Y.}~\bibnamefont {Oreg}}, \bibinfo {author} {\bibfnamefont {C.~M.}\
  \bibnamefont {Marcus}}, \ and\ \bibinfo {author} {\bibfnamefont {M.~H.}\
  \bibnamefont {Freedman}},\ }\href {\doibase 10.1103/PhysRevB.95.235305}
  {\bibfield  {journal} {\bibinfo  {journal} {Phys. Rev. B}\ }\textbf {\bibinfo
  {volume} {95}},\ \bibinfo {pages} {235305} (\bibinfo {year}
  {2017})}\BibitemShut {NoStop}%
\bibitem [{\citenamefont {Plugge}\ \emph {et~al.}(2017)\citenamefont {Plugge},
  \citenamefont {Rasmussen}, \citenamefont {Egger},\ and\ \citenamefont
  {Flensberg}}]{Plugge2017Jan}%
  \BibitemOpen
  \bibfield  {author} {\bibinfo {author} {\bibfnamefont {S.}~\bibnamefont
  {Plugge}}, \bibinfo {author} {\bibfnamefont {A.}~\bibnamefont {Rasmussen}},
  \bibinfo {author} {\bibfnamefont {R.}~\bibnamefont {Egger}}, \ and\ \bibinfo
  {author} {\bibfnamefont {K.}~\bibnamefont {Flensberg}},\ }\href {\doibase
  10.1088/1367-2630/aa54e1} {\bibfield  {journal} {\bibinfo  {journal} {New J.
  Phys.}\ }\textbf {\bibinfo {volume} {19}},\ \bibinfo {pages} {012001}
  (\bibinfo {year} {2017})}\BibitemShut {NoStop}%
\bibitem [{\citenamefont {Litinski}\ \emph {et~al.}(2017)\citenamefont
  {Litinski}, \citenamefont {Kesselring}, \citenamefont {Eisert},\ and\
  \citenamefont {von Oppen}}]{Litinski2017Sep}%
  \BibitemOpen
  \bibfield  {author} {\bibinfo {author} {\bibfnamefont {D.}~\bibnamefont
  {Litinski}}, \bibinfo {author} {\bibfnamefont {M.~S.}\ \bibnamefont
  {Kesselring}}, \bibinfo {author} {\bibfnamefont {J.}~\bibnamefont {Eisert}},
  \ and\ \bibinfo {author} {\bibfnamefont {F.}~\bibnamefont {von Oppen}},\
  }\href {\doibase 10.1103/PhysRevX.7.031048} {\bibfield  {journal} {\bibinfo
  {journal} {Phys. Rev. X}\ }\textbf {\bibinfo {volume} {7}},\ \bibinfo {pages}
  {031048} (\bibinfo {year} {2017})}\BibitemShut {NoStop}%
\bibitem [{\citenamefont {Deng}\ \emph {et~al.}(2012)\citenamefont {Deng},
  \citenamefont {Yu}, \citenamefont {Huang}, \citenamefont {Larsson},
  \citenamefont {Caroff},\ and\ \citenamefont {Xu}}]{Deng2012Dec}%
  \BibitemOpen
  \bibfield  {author} {\bibinfo {author} {\bibfnamefont {M.~T.}\ \bibnamefont
  {Deng}}, \bibinfo {author} {\bibfnamefont {C.~L.}\ \bibnamefont {Yu}},
  \bibinfo {author} {\bibfnamefont {G.~Y.}\ \bibnamefont {Huang}}, \bibinfo
  {author} {\bibfnamefont {M.}~\bibnamefont {Larsson}}, \bibinfo {author}
  {\bibfnamefont {P.}~\bibnamefont {Caroff}}, \ and\ \bibinfo {author}
  {\bibfnamefont {H.~Q.}\ \bibnamefont {Xu}},\ }\href {\doibase
  10.1021/nl303758w} {\bibfield  {journal} {\bibinfo  {journal} {Nano Lett.}\
  }\textbf {\bibinfo {volume} {12}},\ \bibinfo {pages} {6414} (\bibinfo {year}
  {2012})}\BibitemShut {NoStop}%
\bibitem [{\citenamefont {Liu}\ \emph {et~al.}(2012)\citenamefont {Liu},
  \citenamefont {Potter}, \citenamefont {Law},\ and\ \citenamefont
  {Lee}}]{Liu2012Dec}%
  \BibitemOpen
  \bibfield  {author} {\bibinfo {author} {\bibfnamefont {J.}~\bibnamefont
  {Liu}}, \bibinfo {author} {\bibfnamefont {A.~C.}\ \bibnamefont {Potter}},
  \bibinfo {author} {\bibfnamefont {K.~T.}\ \bibnamefont {Law}}, \ and\
  \bibinfo {author} {\bibfnamefont {P.~A.}\ \bibnamefont {Lee}},\ }\href
  {\doibase 10.1103/PhysRevLett.109.267002} {\bibfield  {journal} {\bibinfo
  {journal} {Phys. Rev. Lett.}\ }\textbf {\bibinfo {volume} {109}},\ \bibinfo
  {pages} {267002} (\bibinfo {year} {2012})}\BibitemShut {NoStop}%
\bibitem [{\citenamefont {Kwon}\ \emph {et~al.}(2004)\citenamefont {Kwon},
  \citenamefont {Sengupta},\ and\ \citenamefont {Yakovenko}}]{Kwon2004Feb}%
  \BibitemOpen
  \bibfield  {author} {\bibinfo {author} {\bibfnamefont {H.-J.}\ \bibnamefont
  {Kwon}}, \bibinfo {author} {\bibfnamefont {K.}~\bibnamefont {Sengupta}}, \
  and\ \bibinfo {author} {\bibfnamefont {V.~M.}\ \bibnamefont {Yakovenko}},\
  }\href {\doibase 10.1140/epjb/e2004-00066-4} {\bibfield  {journal} {\bibinfo
  {journal} {Eur. Phys. J. B}\ }\textbf {\bibinfo {volume} {37}},\ \bibinfo
  {pages} {349} (\bibinfo {year} {2004})}\BibitemShut {NoStop}%
\bibitem [{\citenamefont {van Heck}\ \emph {et~al.}(2011)\citenamefont {van
  Heck}, \citenamefont {Hassler}, \citenamefont {Akhmerov},\ and\ \citenamefont
  {Beenakker}}]{vanHeck2011Nov}%
  \BibitemOpen
  \bibfield  {author} {\bibinfo {author} {\bibfnamefont {B.}~\bibnamefont {van
  Heck}}, \bibinfo {author} {\bibfnamefont {F.}~\bibnamefont {Hassler}},
  \bibinfo {author} {\bibfnamefont {A.~R.}\ \bibnamefont {Akhmerov}}, \ and\
  \bibinfo {author} {\bibfnamefont {C.~W.~J.}\ \bibnamefont {Beenakker}},\
  }\href {\doibase 10.1103/PhysRevB.84.180502} {\bibfield  {journal} {\bibinfo
  {journal} {Phys. Rev. B}\ }\textbf {\bibinfo {volume} {84}},\ \bibinfo
  {pages} {180502(R)} (\bibinfo {year} {2011})}\BibitemShut {NoStop}%
\bibitem [{\citenamefont {Law}\ and\ \citenamefont {Lee}(2011)}]{Law2011Aug}%
  \BibitemOpen
  \bibfield  {author} {\bibinfo {author} {\bibfnamefont {K.~T.}\ \bibnamefont
  {Law}}\ and\ \bibinfo {author} {\bibfnamefont {P.~A.}\ \bibnamefont {Lee}},\
  }\href {\doibase 10.1103/PhysRevB.84.081304} {\bibfield  {journal} {\bibinfo
  {journal} {Phys. Rev. B}\ }\textbf {\bibinfo {volume} {84}},\ \bibinfo
  {pages} {081304(R)} (\bibinfo {year} {2011})}\BibitemShut {NoStop}%
\bibitem [{\citenamefont
  {Stefa{\ifmmode\acute{n}\else\'{n}\fi}ski}(2016)}]{Stefanski2016Oct}%
  \BibitemOpen
  \bibfield  {author} {\bibinfo {author} {\bibfnamefont {P.}~\bibnamefont
  {Stefa{\ifmmode\acute{n}\else\'{n}\fi}ski}},\ }\href {\doibase
  10.1088/0953-8984/28/50/505301} {\bibfield  {journal} {\bibinfo  {journal}
  {J. Phys.: Condens. Matter}\ }\textbf {\bibinfo {volume} {28}},\ \bibinfo
  {pages} {505301} (\bibinfo {year} {2016})}\BibitemShut {NoStop}%
\bibitem [{\citenamefont {Laroche}\ \emph {et~al.}(2019)\citenamefont
  {Laroche}, \citenamefont {Bouman}, \citenamefont {van Woerkom}, \citenamefont
  {Proutski}, \citenamefont {Murthy}, \citenamefont {Pikulin}, \citenamefont
  {Nayak}, \citenamefont {van Gulik}, \citenamefont {Nyg{\aa}rd}, \citenamefont
  {Krogstrup}, \citenamefont {Kouwenhoven},\ and\ \citenamefont
  {Geresdi}}]{Laroche2019Jan}%
  \BibitemOpen
  \bibfield  {author} {\bibinfo {author} {\bibfnamefont {D.}~\bibnamefont
  {Laroche}}, \bibinfo {author} {\bibfnamefont {D.}~\bibnamefont {Bouman}},
  \bibinfo {author} {\bibfnamefont {D.~J.}\ \bibnamefont {van Woerkom}},
  \bibinfo {author} {\bibfnamefont {A.}~\bibnamefont {Proutski}}, \bibinfo
  {author} {\bibfnamefont {C.}~\bibnamefont {Murthy}}, \bibinfo {author}
  {\bibfnamefont {D.~I.}\ \bibnamefont {Pikulin}}, \bibinfo {author}
  {\bibfnamefont {C.}~\bibnamefont {Nayak}}, \bibinfo {author} {\bibfnamefont
  {R.~J.~J.}\ \bibnamefont {van Gulik}}, \bibinfo {author} {\bibfnamefont
  {J.}~\bibnamefont {Nyg{\aa}rd}}, \bibinfo {author} {\bibfnamefont
  {P.}~\bibnamefont {Krogstrup}}, \bibinfo {author} {\bibfnamefont {L.~P.}\
  \bibnamefont {Kouwenhoven}}, \ and\ \bibinfo {author} {\bibfnamefont
  {A.}~\bibnamefont {Geresdi}},\ }\href {\doibase 10.1038/s41467-018-08161-2}
  {\bibfield  {journal} {\bibinfo  {journal} {Nat. Commun.}\ }\textbf {\bibinfo
  {volume} {10}},\ \bibinfo {pages} {1} (\bibinfo {year} {2019})}\BibitemShut
  {NoStop}%
\bibitem [{\citenamefont {Hartman}\ \emph {et~al.}(2018)\citenamefont
  {Hartman}, \citenamefont {Olsen}, \citenamefont
  {L{\ifmmode\ddot{u}\else\"{u}\fi}scher}, \citenamefont {Samani},
  \citenamefont {Fallahi}, \citenamefont {Gardner}, \citenamefont {Manfra},\
  and\ \citenamefont {Folk}}]{Hartman2018Aug}%
  \BibitemOpen
  \bibfield  {author} {\bibinfo {author} {\bibfnamefont {N.}~\bibnamefont
  {Hartman}}, \bibinfo {author} {\bibfnamefont {C.}~\bibnamefont {Olsen}},
  \bibinfo {author} {\bibfnamefont {S.}~\bibnamefont
  {L{\ifmmode\ddot{u}\else\"{u}\fi}scher}}, \bibinfo {author} {\bibfnamefont
  {M.}~\bibnamefont {Samani}}, \bibinfo {author} {\bibfnamefont
  {S.}~\bibnamefont {Fallahi}}, \bibinfo {author} {\bibfnamefont {G.~C.}\
  \bibnamefont {Gardner}}, \bibinfo {author} {\bibfnamefont {M.}~\bibnamefont
  {Manfra}}, \ and\ \bibinfo {author} {\bibfnamefont {J.}~\bibnamefont
  {Folk}},\ }\href {\doibase 10.1038/s41567-018-0250-5} {\bibfield  {journal}
  {\bibinfo  {journal} {Nat. Phys.}\ }\textbf {\bibinfo {volume} {14}},\
  \bibinfo {pages} {1083} (\bibinfo {year} {2018})}\BibitemShut {NoStop}%
\bibitem [{\citenamefont {Sela}\ \emph {et~al.}(2019)\citenamefont {Sela},
  \citenamefont {Oreg}, \citenamefont {Plugge}, \citenamefont {Hartman},
  \citenamefont {L{\ifmmode\ddot{u}\else\"{u}\fi}scher},\ and\ \citenamefont
  {Folk}}]{Sela2019Oct}%
  \BibitemOpen
  \bibfield  {author} {\bibinfo {author} {\bibfnamefont {E.}~\bibnamefont
  {Sela}}, \bibinfo {author} {\bibfnamefont {Y.}~\bibnamefont {Oreg}}, \bibinfo
  {author} {\bibfnamefont {S.}~\bibnamefont {Plugge}}, \bibinfo {author}
  {\bibfnamefont {N.}~\bibnamefont {Hartman}}, \bibinfo {author} {\bibfnamefont
  {S.}~\bibnamefont {L{\ifmmode\ddot{u}\else\"{u}\fi}scher}}, \ and\ \bibinfo
  {author} {\bibfnamefont {J.}~\bibnamefont {Folk}},\ }\href {\doibase
  10.1103/PhysRevLett.123.147702} {\bibfield  {journal} {\bibinfo  {journal}
  {Phys. Rev. Lett.}\ }\textbf {\bibinfo {volume} {123}},\ \bibinfo {pages}
  {147702} (\bibinfo {year} {2019})}\BibitemShut {NoStop}%
\bibitem [{\citenamefont {Cooper}\ and\ \citenamefont
  {Stern}(2009)}]{Cooper2009Apr}%
  \BibitemOpen
  \bibfield  {author} {\bibinfo {author} {\bibfnamefont {N.~R.}\ \bibnamefont
  {Cooper}}\ and\ \bibinfo {author} {\bibfnamefont {A.}~\bibnamefont {Stern}},\
  }\href {\doibase 10.1103/PhysRevLett.102.176807} {\bibfield  {journal}
  {\bibinfo  {journal} {Phys. Rev. Lett.}\ }\textbf {\bibinfo {volume} {102}},\
  \bibinfo {pages} {176807} (\bibinfo {year} {2009})}\BibitemShut {NoStop}%
\bibitem [{\citenamefont {Hou}\ \emph {et~al.}(2012)\citenamefont {Hou},
  \citenamefont {Shtengel}, \citenamefont {Refael},\ and\ \citenamefont
  {Goldbart}}]{Hou2012Oct}%
  \BibitemOpen
  \bibfield  {author} {\bibinfo {author} {\bibfnamefont {C.-Y.}\ \bibnamefont
  {Hou}}, \bibinfo {author} {\bibfnamefont {K.}~\bibnamefont {Shtengel}},
  \bibinfo {author} {\bibfnamefont {G.}~\bibnamefont {Refael}}, \ and\ \bibinfo
  {author} {\bibfnamefont {P.~M.}\ \bibnamefont {Goldbart}},\ }\href {\doibase
  10.1088/1367-2630/14/10/105005} {\bibfield  {journal} {\bibinfo  {journal}
  {New J. Phys.}\ }\textbf {\bibinfo {volume} {14}},\ \bibinfo {pages} {105005}
  (\bibinfo {year} {2012})}\BibitemShut {NoStop}%
\bibitem [{\citenamefont {Smirnov}(2015)}]{Smirnov2015Nov}%
  \BibitemOpen
  \bibfield  {author} {\bibinfo {author} {\bibfnamefont {S.}~\bibnamefont
  {Smirnov}},\ }\href {\doibase 10.1103/PhysRevB.92.195312} {\bibfield
  {journal} {\bibinfo  {journal} {Phys. Rev. B}\ }\textbf {\bibinfo {volume}
  {92}},\ \bibinfo {pages} {195312} (\bibinfo {year} {2015})}\BibitemShut
  {NoStop}%
\bibitem [{\citenamefont {Liu}\ and\ \citenamefont
  {Baranger}(2011)}]{Liu2011Nov}%
  \BibitemOpen
  \bibfield  {author} {\bibinfo {author} {\bibfnamefont {D.~E.}\ \bibnamefont
  {Liu}}\ and\ \bibinfo {author} {\bibfnamefont {H.~U.}\ \bibnamefont
  {Baranger}},\ }\href {\doibase 10.1103/PhysRevB.84.201308} {\bibfield
  {journal} {\bibinfo  {journal} {Phys. Rev. B}\ }\textbf {\bibinfo {volume}
  {84}},\ \bibinfo {pages} {201308(R)} (\bibinfo {year} {2011})}\BibitemShut
  {NoStop}%
\bibitem [{\citenamefont {Leijnse}\ and\ \citenamefont
  {Flensberg}(2011)}]{Leijnse2011Oct}%
  \BibitemOpen
  \bibfield  {author} {\bibinfo {author} {\bibfnamefont {M.}~\bibnamefont
  {Leijnse}}\ and\ \bibinfo {author} {\bibfnamefont {K.}~\bibnamefont
  {Flensberg}},\ }\href {\doibase 10.1103/PhysRevB.84.140501} {\bibfield
  {journal} {\bibinfo  {journal} {Phys. Rev. B}\ }\textbf {\bibinfo {volume}
  {84}},\ \bibinfo {pages} {140501(R)} (\bibinfo {year} {2011})}\BibitemShut
  {NoStop}%
\bibitem [{\citenamefont {Cao}\ \emph {et~al.}(2012)\citenamefont {Cao},
  \citenamefont {Wang}, \citenamefont {Xiong}, \citenamefont {Gong},\ and\
  \citenamefont {Li}}]{Cao2012Sep}%
  \BibitemOpen
  \bibfield  {author} {\bibinfo {author} {\bibfnamefont {Y.}~\bibnamefont
  {Cao}}, \bibinfo {author} {\bibfnamefont {P.}~\bibnamefont {Wang}}, \bibinfo
  {author} {\bibfnamefont {G.}~\bibnamefont {Xiong}}, \bibinfo {author}
  {\bibfnamefont {M.}~\bibnamefont {Gong}}, \ and\ \bibinfo {author}
  {\bibfnamefont {X.-Q.}\ \bibnamefont {Li}},\ }\href {\doibase
  10.1103/PhysRevB.86.115311} {\bibfield  {journal} {\bibinfo  {journal} {Phys.
  Rev. B}\ }\textbf {\bibinfo {volume} {86}},\ \bibinfo {pages} {115311}
  (\bibinfo {year} {2012})}\BibitemShut {NoStop}%
\bibitem [{\citenamefont {Lee}\ \emph {et~al.}(2013{\natexlab{b}})\citenamefont
  {Lee}, \citenamefont {Lim},\ and\ \citenamefont
  {L{\ifmmode\acute{o}\else\'{o}\fi}pez}}]{Lee2013Jun}%
  \BibitemOpen
  \bibfield  {author} {\bibinfo {author} {\bibfnamefont {M.}~\bibnamefont
  {Lee}}, \bibinfo {author} {\bibfnamefont {J.~S.}\ \bibnamefont {Lim}}, \ and\
  \bibinfo {author} {\bibfnamefont {R.}~\bibnamefont
  {L{\ifmmode\acute{o}\else\'{o}\fi}pez}},\ }\href {\doibase
  10.1103/PhysRevB.87.241402} {\bibfield  {journal} {\bibinfo  {journal} {Phys.
  Rev. B}\ }\textbf {\bibinfo {volume} {87}},\ \bibinfo {pages} {241402(R)}
  (\bibinfo {year} {2013}{\natexlab{b}})}\BibitemShut {NoStop}%
\bibitem [{\citenamefont {Vernek}\ \emph {et~al.}(2014)\citenamefont {Vernek},
  \citenamefont {Penteado}, \citenamefont {Seridonio},\ and\ \citenamefont
  {Egues}}]{Vernek2014Apr}%
  \BibitemOpen
  \bibfield  {author} {\bibinfo {author} {\bibfnamefont {E.}~\bibnamefont
  {Vernek}}, \bibinfo {author} {\bibfnamefont {P.~H.}\ \bibnamefont
  {Penteado}}, \bibinfo {author} {\bibfnamefont {A.~C.}\ \bibnamefont
  {Seridonio}}, \ and\ \bibinfo {author} {\bibfnamefont {J.~C.}\ \bibnamefont
  {Egues}},\ }\href {\doibase 10.1103/PhysRevB.89.165314} {\bibfield  {journal}
  {\bibinfo  {journal} {Phys. Rev. B}\ }\textbf {\bibinfo {volume} {89}},\
  \bibinfo {pages} {165314} (\bibinfo {year} {2014})}\BibitemShut {NoStop}%
\bibitem [{\citenamefont {Ruiz-Tijerina}\ \emph {et~al.}(2015)\citenamefont
  {Ruiz-Tijerina}, \citenamefont {Vernek}, \citenamefont {Dias~da Silva},\ and\
  \citenamefont {Egues}}]{Ruiz-Tijerina2015Mar}%
  \BibitemOpen
  \bibfield  {author} {\bibinfo {author} {\bibfnamefont {D.~A.}\ \bibnamefont
  {Ruiz-Tijerina}}, \bibinfo {author} {\bibfnamefont {E.}~\bibnamefont
  {Vernek}}, \bibinfo {author} {\bibfnamefont {L.~G. G.~V.}\ \bibnamefont
  {Dias~da Silva}}, \ and\ \bibinfo {author} {\bibfnamefont {J.~C.}\
  \bibnamefont {Egues}},\ }\href {\doibase 10.1103/PhysRevB.91.115435}
  {\bibfield  {journal} {\bibinfo  {journal} {Phys. Rev. B}\ }\textbf {\bibinfo
  {volume} {91}},\ \bibinfo {pages} {115435} (\bibinfo {year}
  {2015})}\BibitemShut {NoStop}%
\bibitem [{\citenamefont {Albrecht}\ \emph {et~al.}(2016)\citenamefont
  {Albrecht}, \citenamefont {Higginbotham}, \citenamefont {Madsen},
  \citenamefont {Kuemmeth}, \citenamefont {Jespersen}, \citenamefont
  {Nyg{\aa}rd}, \citenamefont {Krogstrup},\ and\ \citenamefont
  {Marcus}}]{Albrecht2016Mar}%
  \BibitemOpen
  \bibfield  {author} {\bibinfo {author} {\bibfnamefont {S.~M.}\ \bibnamefont
  {Albrecht}}, \bibinfo {author} {\bibfnamefont {A.~P.}\ \bibnamefont
  {Higginbotham}}, \bibinfo {author} {\bibfnamefont {M.}~\bibnamefont
  {Madsen}}, \bibinfo {author} {\bibfnamefont {F.}~\bibnamefont {Kuemmeth}},
  \bibinfo {author} {\bibfnamefont {T.~S.}\ \bibnamefont {Jespersen}}, \bibinfo
  {author} {\bibfnamefont {J.}~\bibnamefont {Nyg{\aa}rd}}, \bibinfo {author}
  {\bibfnamefont {P.}~\bibnamefont {Krogstrup}}, \ and\ \bibinfo {author}
  {\bibfnamefont {C.~M.}\ \bibnamefont {Marcus}},\ }\href {\doibase
  10.1038/nature17162} {\bibfield  {journal} {\bibinfo  {journal} {Nature}\
  }\textbf {\bibinfo {volume} {531}},\ \bibinfo {pages} {206} (\bibinfo {year}
  {2016})}\BibitemShut {NoStop}%
\bibitem [{\citenamefont {Hoffman}\ \emph {et~al.}(2017)\citenamefont
  {Hoffman}, \citenamefont {Chevallier}, \citenamefont {Loss},\ and\
  \citenamefont {Klinovaja}}]{Hoffman2017Jul}%
  \BibitemOpen
  \bibfield  {author} {\bibinfo {author} {\bibfnamefont {S.}~\bibnamefont
  {Hoffman}}, \bibinfo {author} {\bibfnamefont {D.}~\bibnamefont {Chevallier}},
  \bibinfo {author} {\bibfnamefont {D.}~\bibnamefont {Loss}}, \ and\ \bibinfo
  {author} {\bibfnamefont {J.}~\bibnamefont {Klinovaja}},\ }\href {\doibase
  10.1103/PhysRevB.96.045440} {\bibfield  {journal} {\bibinfo  {journal} {Phys.
  Rev. B}\ }\textbf {\bibinfo {volume} {96}},\ \bibinfo {pages} {045440}
  (\bibinfo {year} {2017})}\BibitemShut {NoStop}%
\bibitem [{\citenamefont {Prada}\ \emph {et~al.}(2017)\citenamefont {Prada},
  \citenamefont {Aguado},\ and\ \citenamefont {San-Jose}}]{Prada2017Aug}%
  \BibitemOpen
  \bibfield  {author} {\bibinfo {author} {\bibfnamefont {E.}~\bibnamefont
  {Prada}}, \bibinfo {author} {\bibfnamefont {R.}~\bibnamefont {Aguado}}, \
  and\ \bibinfo {author} {\bibfnamefont {P.}~\bibnamefont {San-Jose}},\ }\href
  {\doibase 10.1103/PhysRevB.96.085418} {\bibfield  {journal} {\bibinfo
  {journal} {Phys. Rev. B}\ }\textbf {\bibinfo {volume} {96}},\ \bibinfo
  {pages} {085418} (\bibinfo {year} {2017})}\BibitemShut {NoStop}%
\bibitem [{\citenamefont {Clarke}(2017)}]{Clarke2017Nov}%
  \BibitemOpen
  \bibfield  {author} {\bibinfo {author} {\bibfnamefont {D.~J.}\ \bibnamefont
  {Clarke}},\ }\href {\doibase 10.1103/PhysRevB.96.201109} {\bibfield
  {journal} {\bibinfo  {journal} {Phys. Rev. B}\ }\textbf {\bibinfo {volume}
  {96}},\ \bibinfo {pages} {201109(R)} (\bibinfo {year} {2017})}\BibitemShut
  {NoStop}%
\bibitem [{\citenamefont {Chevallier}\ \emph {et~al.}(2018)\citenamefont
  {Chevallier}, \citenamefont {Szumniak}, \citenamefont {Hoffman},
  \citenamefont {Loss},\ and\ \citenamefont {Klinovaja}}]{Chevallier2018Jan}%
  \BibitemOpen
  \bibfield  {author} {\bibinfo {author} {\bibfnamefont {D.}~\bibnamefont
  {Chevallier}}, \bibinfo {author} {\bibfnamefont {P.}~\bibnamefont
  {Szumniak}}, \bibinfo {author} {\bibfnamefont {S.}~\bibnamefont {Hoffman}},
  \bibinfo {author} {\bibfnamefont {D.}~\bibnamefont {Loss}}, \ and\ \bibinfo
  {author} {\bibfnamefont {J.}~\bibnamefont {Klinovaja}},\ }\href {\doibase
  10.1103/PhysRevB.97.045404} {\bibfield  {journal} {\bibinfo  {journal} {Phys.
  Rev. B}\ }\textbf {\bibinfo {volume} {97}},\ \bibinfo {pages} {045404}
  (\bibinfo {year} {2018})}\BibitemShut {NoStop}%
\bibitem [{\citenamefont {Deng}\ \emph {et~al.}(2018)\citenamefont {Deng},
  \citenamefont {Vaitiek{\ifmmode\dot{e}\else\.{e}\fi}nas}, \citenamefont
  {Prada}, \citenamefont {San-Jose}, \citenamefont {Nyg{\aa}rd}, \citenamefont
  {Krogstrup}, \citenamefont {Aguado},\ and\ \citenamefont
  {Marcus}}]{Deng2018Aug}%
  \BibitemOpen
  \bibfield  {author} {\bibinfo {author} {\bibfnamefont {M.-T.}\ \bibnamefont
  {Deng}}, \bibinfo {author} {\bibfnamefont {S.}~\bibnamefont
  {Vaitiek{\ifmmode\dot{e}\else\.{e}\fi}nas}}, \bibinfo {author} {\bibfnamefont
  {E.}~\bibnamefont {Prada}}, \bibinfo {author} {\bibfnamefont
  {P.}~\bibnamefont {San-Jose}}, \bibinfo {author} {\bibfnamefont
  {J.}~\bibnamefont {Nyg{\aa}rd}}, \bibinfo {author} {\bibfnamefont
  {P.}~\bibnamefont {Krogstrup}}, \bibinfo {author} {\bibfnamefont
  {R.}~\bibnamefont {Aguado}}, \ and\ \bibinfo {author} {\bibfnamefont {C.~M.}\
  \bibnamefont {Marcus}},\ }\href {\doibase 10.1103/PhysRevB.98.085125}
  {\bibfield  {journal} {\bibinfo  {journal} {Phys. Rev. B}\ }\textbf {\bibinfo
  {volume} {98}},\ \bibinfo {pages} {085125} (\bibinfo {year}
  {2018})}\BibitemShut {NoStop}%
\bibitem [{\citenamefont {Terhal}\ \emph {et~al.}(2012)\citenamefont {Terhal},
  \citenamefont {Hassler},\ and\ \citenamefont {DiVincenzo}}]{Terhal2012Jun}%
  \BibitemOpen
  \bibfield  {author} {\bibinfo {author} {\bibfnamefont {B.~M.}\ \bibnamefont
  {Terhal}}, \bibinfo {author} {\bibfnamefont {F.}~\bibnamefont {Hassler}}, \
  and\ \bibinfo {author} {\bibfnamefont {D.~P.}\ \bibnamefont {DiVincenzo}},\
  }\href {\doibase 10.1103/PhysRevLett.108.260504} {\bibfield  {journal}
  {\bibinfo  {journal} {Phys. Rev. Lett.}\ }\textbf {\bibinfo {volume} {108}},\
  \bibinfo {pages} {260504} (\bibinfo {year} {2012})}\BibitemShut {NoStop}%
\bibitem [{\citenamefont {Sau}\ and\ \citenamefont
  {Das~Sarma}(2012)}]{Sau2012Jul}%
  \BibitemOpen
  \bibfield  {author} {\bibinfo {author} {\bibfnamefont {J.~D.}\ \bibnamefont
  {Sau}}\ and\ \bibinfo {author} {\bibfnamefont {S.}~\bibnamefont
  {Das~Sarma}},\ }\href {\doibase 10.1038/ncomms1966} {\bibfield  {journal}
  {\bibinfo  {journal} {Nat. Commun.}\ }\textbf {\bibinfo {volume} {3}},\
  \bibinfo {pages} {964} (\bibinfo {year} {2012})}\BibitemShut {NoStop}%
\bibitem [{\citenamefont {Cayao}\ \emph {et~al.}(2018)\citenamefont {Cayao},
  \citenamefont {Black-Schaffer}, \citenamefont {Prada},\ and\ \citenamefont
  {Aguado}}]{Cayao2018May}%
  \BibitemOpen
  \bibfield  {author} {\bibinfo {author} {\bibfnamefont {J.}~\bibnamefont
  {Cayao}}, \bibinfo {author} {\bibfnamefont {A.~M.}\ \bibnamefont
  {Black-Schaffer}}, \bibinfo {author} {\bibfnamefont {E.}~\bibnamefont
  {Prada}}, \ and\ \bibinfo {author} {\bibfnamefont {R.}~\bibnamefont
  {Aguado}},\ }\href {\doibase 10.3762/bjnano.9.127} {\bibfield  {journal}
  {\bibinfo  {journal} {Beilstein J. Nanotechnol.}\ }\textbf {\bibinfo {volume}
  {9}},\ \bibinfo {pages} {1339} (\bibinfo {year} {2018})}\BibitemShut
  {NoStop}%
\bibitem [{\citenamefont {O{'}Farrell}\ \emph {et~al.}(2018)\citenamefont
  {O{'}Farrell}, \citenamefont {Drachmann}, \citenamefont {Hell}, \citenamefont
  {Fornieri}, \citenamefont {Whiticar}, \citenamefont {Hansen}, \citenamefont
  {Gronin}, \citenamefont {Gardner}, \citenamefont {Thomas}, \citenamefont
  {Manfra}, \citenamefont {Flensberg}, \citenamefont {Marcus},\ and\
  \citenamefont {Nichele}}]{O'Farrell2018Dec}%
  \BibitemOpen
  \bibfield  {author} {\bibinfo {author} {\bibfnamefont {E.~C.~T.}\
  \bibnamefont {O{'}Farrell}}, \bibinfo {author} {\bibfnamefont {A.~C.~C.}\
  \bibnamefont {Drachmann}}, \bibinfo {author} {\bibfnamefont {M.}~\bibnamefont
  {Hell}}, \bibinfo {author} {\bibfnamefont {A.}~\bibnamefont {Fornieri}},
  \bibinfo {author} {\bibfnamefont {A.~M.}\ \bibnamefont {Whiticar}}, \bibinfo
  {author} {\bibfnamefont {E.~B.}\ \bibnamefont {Hansen}}, \bibinfo {author}
  {\bibfnamefont {S.}~\bibnamefont {Gronin}}, \bibinfo {author} {\bibfnamefont
  {G.~C.}\ \bibnamefont {Gardner}}, \bibinfo {author} {\bibfnamefont
  {C.}~\bibnamefont {Thomas}}, \bibinfo {author} {\bibfnamefont {M.~J.}\
  \bibnamefont {Manfra}}, \bibinfo {author} {\bibfnamefont {K.}~\bibnamefont
  {Flensberg}}, \bibinfo {author} {\bibfnamefont {C.~M.}\ \bibnamefont
  {Marcus}}, \ and\ \bibinfo {author} {\bibfnamefont {F.}~\bibnamefont
  {Nichele}},\ }\href {\doibase 10.1103/PhysRevLett.121.256803} {\bibfield
  {journal} {\bibinfo  {journal} {Phys. Rev. Lett.}\ }\textbf {\bibinfo
  {volume} {121}},\ \bibinfo {pages} {256803} (\bibinfo {year}
  {2018})}\BibitemShut {NoStop}%
\bibitem [{\citenamefont {Ricco}\ \emph {et~al.}(2019)\citenamefont {Ricco},
  \citenamefont {de~Souza}, \citenamefont {Figueira}, \citenamefont {Shelykh},\
  and\ \citenamefont {Seridonio}}]{Ricco2019Apr}%
  \BibitemOpen
  \bibfield  {author} {\bibinfo {author} {\bibfnamefont {L.~S.}\ \bibnamefont
  {Ricco}}, \bibinfo {author} {\bibfnamefont {M.}~\bibnamefont {de~Souza}},
  \bibinfo {author} {\bibfnamefont {M.~S.}\ \bibnamefont {Figueira}}, \bibinfo
  {author} {\bibfnamefont {I.~A.}\ \bibnamefont {Shelykh}}, \ and\ \bibinfo
  {author} {\bibfnamefont {A.~C.}\ \bibnamefont {Seridonio}},\ }\href {\doibase
  10.1103/PhysRevB.99.155159} {\bibfield  {journal} {\bibinfo  {journal} {Phys.
  Rev. B}\ }\textbf {\bibinfo {volume} {99}},\ \bibinfo {pages} {155159}
  (\bibinfo {year} {2019})}\BibitemShut {NoStop}%
\bibitem [{\citenamefont {Yavilberg}\ \emph {et~al.}(2019)\citenamefont
  {Yavilberg}, \citenamefont {Ginossar},\ and\ \citenamefont
  {Grosfeld}}]{Yavilberg2019Dec}%
  \BibitemOpen
  \bibfield  {author} {\bibinfo {author} {\bibfnamefont {K.}~\bibnamefont
  {Yavilberg}}, \bibinfo {author} {\bibfnamefont {E.}~\bibnamefont {Ginossar}},
  \ and\ \bibinfo {author} {\bibfnamefont {E.}~\bibnamefont {Grosfeld}},\
  }\href {\doibase 10.1103/PhysRevB.100.241408} {\bibfield  {journal} {\bibinfo
   {journal} {Phys. Rev. B}\ }\textbf {\bibinfo {volume} {100}},\ \bibinfo
  {pages} {241408(R)} (\bibinfo {year} {2019})}\BibitemShut {NoStop}%
\bibitem [{\citenamefont {Camjayi}\ \emph {et~al.}(2017)\citenamefont
  {Camjayi}, \citenamefont {Arrachea}, \citenamefont {Aligia},\ and\
  \citenamefont {von Oppen}}]{Camjayi2017Jul}%
  \BibitemOpen
  \bibfield  {author} {\bibinfo {author} {\bibfnamefont {A.}~\bibnamefont
  {Camjayi}}, \bibinfo {author} {\bibfnamefont {L.}~\bibnamefont {Arrachea}},
  \bibinfo {author} {\bibfnamefont {A.}~\bibnamefont {Aligia}}, \ and\ \bibinfo
  {author} {\bibfnamefont {F.}~\bibnamefont {von Oppen}},\ }\href {\doibase
  10.1103/PhysRevLett.119.046801} {\bibfield  {journal} {\bibinfo  {journal}
  {Phys. Rev. Lett.}\ }\textbf {\bibinfo {volume} {119}},\ \bibinfo {pages}
  {046801} (\bibinfo {year} {2017})}\BibitemShut {NoStop}%
\bibitem [{\citenamefont {Cayao}\ \emph {et~al.}(2017)\citenamefont {Cayao},
  \citenamefont {San-Jose}, \citenamefont {Black-Schaffer}, \citenamefont
  {Aguado},\ and\ \citenamefont {Prada}}]{Cayao2017Nov}%
  \BibitemOpen
  \bibfield  {author} {\bibinfo {author} {\bibfnamefont {J.}~\bibnamefont
  {Cayao}}, \bibinfo {author} {\bibfnamefont {P.}~\bibnamefont {San-Jose}},
  \bibinfo {author} {\bibfnamefont {A.~M.}\ \bibnamefont {Black-Schaffer}},
  \bibinfo {author} {\bibfnamefont {R.}~\bibnamefont {Aguado}}, \ and\ \bibinfo
  {author} {\bibfnamefont {E.}~\bibnamefont {Prada}},\ }\href {\doibase
  10.1103/PhysRevB.96.205425} {\bibfield  {journal} {\bibinfo  {journal} {Phys.
  Rev. B}\ }\textbf {\bibinfo {volume} {96}},\ \bibinfo {pages} {205425}
  (\bibinfo {year} {2017})}\BibitemShut {NoStop}%
\bibitem [{\citenamefont {Schrade}\ and\ \citenamefont
  {Fu}(2018)}]{Schrade2018Sep}%
  \BibitemOpen
  \bibfield  {author} {\bibinfo {author} {\bibfnamefont {C.}~\bibnamefont
  {Schrade}}\ and\ \bibinfo {author} {\bibfnamefont {L.}~\bibnamefont {Fu}},\
  }\href {https://arxiv.org/abs/1809.06370} {\bibfield  {journal} {\bibinfo
  {journal} {arXiv}\ } (\bibinfo {year} {2018})},\ \Eprint
  {http://arxiv.org/abs/1809.06370} {1809.06370} \BibitemShut {NoStop}%
\bibitem [{\citenamefont {Ridderbos}\ \emph {et~al.}(2018)\citenamefont
  {Ridderbos}, \citenamefont {Brauns}, \citenamefont {Shen}, \citenamefont
  {de~Vries}, \citenamefont {Li}, \citenamefont {Bakkers}, \citenamefont
  {Brinkman},\ and\ \citenamefont {Zwanenburg}}]{Ridderbos2018Nov}%
  \BibitemOpen
  \bibfield  {author} {\bibinfo {author} {\bibfnamefont {J.}~\bibnamefont
  {Ridderbos}}, \bibinfo {author} {\bibfnamefont {M.}~\bibnamefont {Brauns}},
  \bibinfo {author} {\bibfnamefont {J.}~\bibnamefont {Shen}}, \bibinfo {author}
  {\bibfnamefont {F.~K.}\ \bibnamefont {de~Vries}}, \bibinfo {author}
  {\bibfnamefont {A.}~\bibnamefont {Li}}, \bibinfo {author} {\bibfnamefont
  {E.~P. A.~M.}\ \bibnamefont {Bakkers}}, \bibinfo {author} {\bibfnamefont
  {A.}~\bibnamefont {Brinkman}}, \ and\ \bibinfo {author} {\bibfnamefont
  {F.~A.}\ \bibnamefont {Zwanenburg}},\ }\href {\doibase
  10.1002/adma.201802257} {\bibfield  {journal} {\bibinfo  {journal} {Adv.
  Mater.}\ }\textbf {\bibinfo {volume} {30}},\ \bibinfo {pages} {1802257}
  (\bibinfo {year} {2018})}\BibitemShut {NoStop}%
\bibitem [{\citenamefont {Awoga}\ \emph {et~al.}(2019)\citenamefont {Awoga},
  \citenamefont {Cayao},\ and\ \citenamefont {Black-Schaffer}}]{Awoga2019Sep}%
  \BibitemOpen
  \bibfield  {author} {\bibinfo {author} {\bibfnamefont {O.~A.}\ \bibnamefont
  {Awoga}}, \bibinfo {author} {\bibfnamefont {J.}~\bibnamefont {Cayao}}, \ and\
  \bibinfo {author} {\bibfnamefont {A.~M.}\ \bibnamefont {Black-Schaffer}},\
  }\href {\doibase 10.1103/PhysRevLett.123.117001} {\bibfield  {journal}
  {\bibinfo  {journal} {Phys. Rev. Lett.}\ }\textbf {\bibinfo {volume} {123}},\
  \bibinfo {pages} {117001} (\bibinfo {year} {2019})}\BibitemShut {NoStop}%
\bibitem [{\citenamefont {Spivak}\ and\ \citenamefont
  {Kivelson}(1991)}]{Spivak1991Feb}%
  \BibitemOpen
  \bibfield  {author} {\bibinfo {author} {\bibfnamefont {B.~I.}\ \bibnamefont
  {Spivak}}\ and\ \bibinfo {author} {\bibfnamefont {S.~A.}\ \bibnamefont
  {Kivelson}},\ }\href {\doibase 10.1103/PhysRevB.43.3740} {\bibfield
  {journal} {\bibinfo  {journal} {Phys. Rev. B}\ }\textbf {\bibinfo {volume}
  {43}},\ \bibinfo {pages} {3740} (\bibinfo {year} {1991})}\BibitemShut
  {NoStop}%
\bibitem [{\citenamefont {Rozhkov}\ \emph {et~al.}(2001)\citenamefont
  {Rozhkov}, \citenamefont {Arovas},\ and\ \citenamefont
  {Guinea}}]{Rozhkov2001Nov}%
  \BibitemOpen
  \bibfield  {author} {\bibinfo {author} {\bibfnamefont {A.~V.}\ \bibnamefont
  {Rozhkov}}, \bibinfo {author} {\bibfnamefont {D.~P.}\ \bibnamefont {Arovas}},
  \ and\ \bibinfo {author} {\bibfnamefont {F.}~\bibnamefont {Guinea}},\ }\href
  {\doibase 10.1103/PhysRevB.64.233301} {\bibfield  {journal} {\bibinfo
  {journal} {Phys. Rev. B}\ }\textbf {\bibinfo {volume} {64}},\ \bibinfo
  {pages} {233301} (\bibinfo {year} {2001})}\BibitemShut {NoStop}%
\bibitem [{\citenamefont {van Dam}\ \emph {et~al.}(2006)\citenamefont {van
  Dam}, \citenamefont {Nazarov}, \citenamefont {Bakkers}, \citenamefont
  {De~Franceschi},\ and\ \citenamefont {Kouwenhoven}}]{vanDam2006Aug}%
  \BibitemOpen
  \bibfield  {author} {\bibinfo {author} {\bibfnamefont {J.~A.}\ \bibnamefont
  {van Dam}}, \bibinfo {author} {\bibfnamefont {Y.~V.}\ \bibnamefont
  {Nazarov}}, \bibinfo {author} {\bibfnamefont {E.~P. A.~M.}\ \bibnamefont
  {Bakkers}}, \bibinfo {author} {\bibfnamefont {S.}~\bibnamefont
  {De~Franceschi}}, \ and\ \bibinfo {author} {\bibfnamefont {L.~P.}\
  \bibnamefont {Kouwenhoven}},\ }\href {\doibase 10.1038/nature05018}
  {\bibfield  {journal} {\bibinfo  {journal} {Nature}\ }\textbf {\bibinfo
  {volume} {442}},\ \bibinfo {pages} {667} (\bibinfo {year}
  {2006})}\BibitemShut {NoStop}%
\bibitem [{\citenamefont {Hartree}(1928)}]{Hartree1928Jan}%
  \BibitemOpen
  \bibfield  {author} {\bibinfo {author} {\bibfnamefont {D.~R.}\ \bibnamefont
  {Hartree}},\ }\href {\doibase 10.1017/S0305004100011919} {\bibfield
  {journal} {\bibinfo  {journal} {Math. Proc. Cambridge Philos. Soc.}\ }\textbf
  {\bibinfo {volume} {24}},\ \bibinfo {pages} {89} (\bibinfo {year}
  {1928})}\BibitemShut {NoStop}%
\bibitem [{\citenamefont {Pillet}\ \emph {et~al.}(2010)\citenamefont {Pillet},
  \citenamefont {Quay}, \citenamefont {Morfin}, \citenamefont {Bena},
  \citenamefont {Yeyati},\ and\ \citenamefont {Joyez}}]{Pillet2010Nov}%
  \BibitemOpen
  \bibfield  {author} {\bibinfo {author} {\bibfnamefont {J.-D.}\ \bibnamefont
  {Pillet}}, \bibinfo {author} {\bibfnamefont {C.~H.~L.}\ \bibnamefont {Quay}},
  \bibinfo {author} {\bibfnamefont {P.}~\bibnamefont {Morfin}}, \bibinfo
  {author} {\bibfnamefont {C.}~\bibnamefont {Bena}}, \bibinfo {author}
  {\bibfnamefont {A.~L.}\ \bibnamefont {Yeyati}}, \ and\ \bibinfo {author}
  {\bibfnamefont {P.}~\bibnamefont {Joyez}},\ }\href {\doibase
  10.1038/nphys1811} {\bibfield  {journal} {\bibinfo  {journal} {Nat. Phys.}\
  }\textbf {\bibinfo {volume} {6}},\ \bibinfo {pages} {965} (\bibinfo {year}
  {2010})}\BibitemShut {NoStop}%
\bibitem [{\citenamefont {Chang}\ \emph {et~al.}(2013)\citenamefont {Chang},
  \citenamefont {Manucharyan}, \citenamefont {Jespersen}, \citenamefont
  {Nyg{\aa}rd},\ and\ \citenamefont {Marcus}}]{Chang2013May}%
  \BibitemOpen
  \bibfield  {author} {\bibinfo {author} {\bibfnamefont {W.}~\bibnamefont
  {Chang}}, \bibinfo {author} {\bibfnamefont {V.~E.}\ \bibnamefont
  {Manucharyan}}, \bibinfo {author} {\bibfnamefont {T.~S.}\ \bibnamefont
  {Jespersen}}, \bibinfo {author} {\bibfnamefont {J.}~\bibnamefont
  {Nyg{\aa}rd}}, \ and\ \bibinfo {author} {\bibfnamefont {C.~M.}\ \bibnamefont
  {Marcus}},\ }\href {\doibase 10.1103/PhysRevLett.110.217005} {\bibfield
  {journal} {\bibinfo  {journal} {Phys. Rev. Lett.}\ }\textbf {\bibinfo
  {volume} {110}},\ \bibinfo {pages} {217005} (\bibinfo {year}
  {2013})}\BibitemShut {NoStop}%
\bibitem [{\citenamefont {Whiticar}\ \emph {et~al.}(2019)\citenamefont
  {Whiticar}, \citenamefont {Fornieri},\ and\ \citenamefont
  {Marcus}}]{Whiticar2019}%
  \BibitemOpen
  \bibfield  {author} {\bibinfo {author} {\bibfnamefont {A.~M.}\ \bibnamefont
  {Whiticar}}, \bibinfo {author} {\bibfnamefont {A.}~\bibnamefont {Fornieri}},
  \ and\ \bibinfo {author} {\bibfnamefont {C.~M.}\ \bibnamefont {Marcus}},\
  }\href@noop {} {} (\bibinfo {year} {2019}),\ \bibinfo {note} {private
  communication}\BibitemShut {NoStop}%
\bibitem [{\citenamefont {Mahaux}\ and\ \citenamefont
  {Weidenm{\ifmmode\ddot{u}\else\"{u}\fi}ller}(1968)}]{Mahaux1968Jun}%
  \BibitemOpen
  \bibfield  {author} {\bibinfo {author} {\bibfnamefont {C.}~\bibnamefont
  {Mahaux}}\ and\ \bibinfo {author} {\bibfnamefont {H.~A.}\ \bibnamefont
  {Weidenm{\ifmmode\ddot{u}\else\"{u}\fi}ller}},\ }\href {\doibase
  10.1103/PhysRev.170.847} {\bibfield  {journal} {\bibinfo  {journal} {Phys.
  Rev.}\ }\textbf {\bibinfo {volume} {170}},\ \bibinfo {pages} {847} (\bibinfo
  {year} {1968})}\BibitemShut {NoStop}%
\bibitem [{\citenamefont {Yu}(1965)}]{Yu1965}%
  \BibitemOpen
  \bibfield  {author} {\bibinfo {author} {\bibfnamefont {L.}~\bibnamefont
  {Yu}},\ }\href {\doibase 10.7498/aps.21.75} {\bibfield  {journal} {\bibinfo
  {journal} {Acta Phys. Sin.}\ }\textbf {\bibinfo {volume} {114}},\ \bibinfo
  {pages} {75} (\bibinfo {year} {1965})}\BibitemShut {NoStop}%
\bibitem [{\citenamefont {Shiba}(1968)}]{Shiba1968Sep}%
  \BibitemOpen
  \bibfield  {author} {\bibinfo {author} {\bibfnamefont {H.}~\bibnamefont
  {Shiba}},\ }\href {\doibase 10.1143/PTP.40.435} {\bibfield  {journal}
  {\bibinfo  {journal} {Prog. Theor. Phys.}\ }\textbf {\bibinfo {volume}
  {40}},\ \bibinfo {pages} {435} (\bibinfo {year} {1968})}\BibitemShut
  {NoStop}%
\bibitem [{\citenamefont {Rusinov}(1969)}]{Rusinov1969}%
  \BibitemOpen
  \bibfield  {author} {\bibinfo {author} {\bibfnamefont {A.~I.}\ \bibnamefont
  {Rusinov}},\ }\href
  {http://www.jetpletters.ac.ru/ps/1658/article_25295.shtml} {\bibfield
  {journal} {\bibinfo  {journal} {JETP Lett.}\ }\textbf {\bibinfo {volume}
  {9}},\ \bibinfo {pages} {85} (\bibinfo {year} {1969})}\BibitemShut {NoStop}%
\bibitem [{\citenamefont {Novotn{\ifmmode\acute{y}\else\'{y}\fi}}\ \emph
  {et~al.}(2005)\citenamefont {Novotn{\ifmmode\acute{y}\else\'{y}\fi}},
  \citenamefont {Rossini},\ and\ \citenamefont {Flensberg}}]{Novotny2005Dec}%
  \BibitemOpen
  \bibfield  {author} {\bibinfo {author} {\bibfnamefont {T.}~\bibnamefont
  {Novotn{\ifmmode\acute{y}\else\'{y}\fi}}}, \bibinfo {author} {\bibfnamefont
  {A.}~\bibnamefont {Rossini}}, \ and\ \bibinfo {author} {\bibfnamefont
  {K.}~\bibnamefont {Flensberg}},\ }\href {\doibase 10.1103/PhysRevB.72.224502}
  {\bibfield  {journal} {\bibinfo  {journal} {Phys. Rev. B}\ }\textbf {\bibinfo
  {volume} {72}},\ \bibinfo {pages} {224502} (\bibinfo {year}
  {2005})}\BibitemShut {NoStop}%
\end{thebibliography}
\end{document}